\documentclass[aps,prl,reprint,groupedaddress,nofootinbib]{revtex4-1}
\usepackage{hyperref}
\usepackage{amsmath}
 \usepackage{multirow}
\usepackage{array}
\newcolumntype{L}[1]{>{\raggedright\let\newline\\\arraybacksslash\hspace{0pt}}m{#1}}
\newcolumntype{C}[1]{>{\centering\let\newline\\\arraybackslash\hspace{0pt}}m{#1}}
\newcolumntype{R}[1]{>{\raggedleft\let\newline\\\arraybackslash\hspace{0pt}}m{#1}}
\usepackage{float}
\usepackage{graphicx}
\usepackage{epsfig}
\usepackage{psfrag}
\usepackage{adjustbox} 
\usepackage{color}
\usepackage{slashed}
\usepackage{appendix}
\usepackage{csquotes}
\usepackage{subcaption}
\usepackage{amsfonts}
\usepackage{amssymb}
\usepackage{soul}
\usepackage{tikz}
\usepackage{tikz}
\usetikzlibrary{positioning,arrows}
\usetikzlibrary{decorations.pathmorphing}
\usetikzlibrary{decorations.markings}

\newcommand*{\be}{\begin{equation}}
\newcommand*{\ee}{\end{equation}}
\newcommand*{\bea}{\begin{eqnarray}}
\newcommand*{\eea}{\end{eqnarray}}

\newcommand{\comment}[1]{}

\newcommand{\cref}[1]{Chapter~\ref{c.#1}}

\def\beq{\begin{equation}}
\def\eeq{\end{equation}}
\def\bea{\begin{eqnarray}}
\def\eea{\end{eqnarray}}
\def\ba{\begin{array}}
\def\ea{\end{array}}
\def\bi{\begin{itemize}}
\def\ei{\end{itemize}}
\def\be{\begin{enumerate}}
\def\ee{\end{enumerate}}
\def\bc{\begin{center}}
\def\ec{\end{center}}
\def\bt{\begin{table}}
\def\et{\end{table}}
\def\btb{\begin{tabular}}
\def\etb{\end{tabular}}

\definecolor{dgreen}{rgb}{0.0, 0.5, 0.0}

\def\lsim{\raise0.3ex\hbox{$\;<$\kern-0.75em\raise-1.1ex\hbox{$\sim\;$}}}
\def\gsim{\raise0.3ex\hbox{$\;>$\kern-0.75em\raise-1.1ex\hbox{$\sim\;$}}}
	
\usepackage{cancel}
 \usepackage[normalem]{ulem}
\usepackage{color}
\def\comment#1{\textcolor{blue}{\large(\it{#1})}}

\def\lapp{\mathrel{\rlap{\raise.5ex\hbox{$<$}}
                    {\lower.5ex\hbox{$\sim$}}}}
\def\gapp{\mathrel{\rlap{\raise.5ex\hbox{$>$}}
                    {\lower.5ex\hbox{$\sim$}}}}
\begin{document}

\title{{Multi-boson splashes at future colliders from electroweak compositeness}}

\author{G. Cacciapaglia$^{1,2}$}
\email{g.cacciapaglia@ipnl.in2p3.fr}

\author{A. Deandrea$^{3,4}$}
\email{deandrea@ip2i.in2p3.fr}

\author{A.M.  Iyer$^{5}$}
\email{iyerabhishek@physics.iitd.ac.in}

\author{S.~Kulkarni$^{6}$}
\email{suchita.kulkarni@uni-graz.at}

\author{A.K Singh$^{5}$}
\email{Abhishek.Kumar.Singh@physics.iitd.ac.in}

\affiliation{
$^1$Laboratoire de Physique Th\'eorique et Hautes \'Energies (LPTHE), UMR 7589,
Sorbonne Universit\'e \& CNRS, 4 place Jussieu, 75252 Paris Cedex 05, France\\
$^2$Quantum Theory Center (QTC) \& D-IAS, Southern Denmark Univ., Campusvej 55, 5230 Odense M, Denmark\\
$^3$
Universit\'e Claude Bernard Lyon 1, CNRS/IN2P3, IP2I UMR 5822, F-69100 Villeurbanne, France\\
$^4$ Department of Physics, University of Johannesburg,
Box 524, Auckland Park 2006, South Africa\\
$^5$ Department of Physics, Indian Institute of Technology Delhi, New Delhi-110016, India\\
$^6$Institute of Physics, NAWI Graz, University of Graz, Universitatsplatz 5, A-8010 Graz, Austria 
}

\begin{abstract}
We propose a new collider signature for the composite origin of the electroweak symmetry breaking of the standard model. The Higgs sector consists of new fundamental fermions (hyper-quarks), which confine at a hadronization scale $\Lambda_{HC} \sim$ few TeV. At energies above $\Lambda_{HC}$, the Drell-Yan production of the hyper-quarks leads to the production of a few electroweak bosons, in analogy with hadron production in QCD at $e^+e^- \to q\bar{q}$ around a few GeV. We show that this regime can be probed at future colliders, namely the proposed 100 TeV hadron collider (FCC-hh) and a 10 TeV muon collider. Together with the direct discovery of electroweak resonances, the multi electroweak boson signature provides a smoking gun for Higgs compositeness.
\end{abstract}

%\pacs{73.21.Hb, 73.21.La, 73.50.Bk}
\maketitle

The origin of the electroweak (EW) symmetry breaking in the Standard Model (SM) remains unresolved. %an unanswered question.  
A class of models that aims at providing a dynamical explanation poses the Higgs boson as a bound state within a new composite sector akin to QCD, where spontaneous symmetry breaking arises. A scaled-up version of QCD, dubbed Technicolor (TC), was the first concrete realization~\cite{Susskind:1978ms, FARHI1981277}. A generic TC-like model consists of a new fundamental fermionic sector (made of hyper-quarks, $q_{HC}$), charged under the EW group and a new confining hyper-color group $G_{HC}$. Similarly to QCD, the new dynamics is characterized by a new physics scale $\Lambda_{HC} \sim \mathcal{O}(\mbox{few})$~TeV, where $q_{HC}$ bound states are present. In this setup, the longitudinal polarization of the EW bosons and the Higgs emerge as light resonances below $\Lambda_{HC}$.

If the $q_{HC}$ condensate breaks the EW symmetry, then one would naturally expect $\Lambda_{HC} \sim 1 - 3$~TeV. There are models where the condensate also features an EW-preserving direction, and the EW symmetry breaking is due to misalignment \cite{Kaplan:1983fs}, hence allowing for larger $\Lambda_{HC}$ up to tens of TeV. In those models, the Higgs boson emerges as a pseudo-Goldstone, while a TC-like direction can always be reached continuously \cite{Cacciapaglia_2014}. The scale $\Lambda_{HC}$ serves as an indicator beyond which the composite theories can be treated in terms of the fundamental hyper-quarks and hyper-gluons and beyond the chiral approximation.
Investigations for compositeness at past and current colliders, such as the Large Electron-Positron (LEP) and the Large Hadron (LHC) colliders, have been focused either on precision measurements in the EW sector (gauge and Higgs couplings) or on searches for individual resonances. This strategy was mainly due to the limited energies in the individual collision, reaching up to a few TeV at partonic level.  This work sketches out a novel strategy for the exploration of TC-like theories at future higher energy colliders, where large multiplicities of composite resonances are produced. %This follows from the fact that for a fixed $\Lambda_{\rm HC}$ of the TC-like theory, a higher centre of mass (com) energy may yield more exotic signatures. 

The exploration of strongly-interacting theories at colliders can be classified into regimes that depend on the available collision energy, as parametrized by partonic Mandelstam variable $\hat{s}$. For $\sqrt{\hat{s}} < \Lambda_{HC}$, chiral expansion is used to study the direct production of a small (one or two) number of light resonances.
For $\sqrt{\hat{s}} > \Lambda_{HC}$, instead, the production of resonances can be parametrized in terms of perturbative matrix elements involving hyper-quarks and hyper-gluons.
Hence, irrespective of the theory, whether QCD or TC-like, the corresponding regimes can be conveniently  quantified by the ratio
$\sqrt{\hat s}/\Lambda$.
In the case of QCD, the inclusive measurement for $e^+e^-\rightarrow$ multi-mesons is consistent with the evaluations from the perturbative matrix element $\mathcal{M}(e^+e^-\rightarrow q\bar q )$ at energies above $\sim 2$~GeV. The mesons are formed from the outgoing quarks by a process of fragmentation and hadronization  \cite{ARTRU197493,Chun:1992qs}.
Depending on the number of produced mesons, the value of $\sqrt{\hat s}/\Lambda_{QCD}$ can further classify the perturbative regime into two broad categories: 
\begin{itemize}
    \item[i)] $\sqrt{\hat s}/\Lambda_{QCD}\sim \mathcal{O}(1 - 10)$ leads to the production of a small number of mesons (process $2\rightarrow \mathit {few}$). In this regime, data is described in terms of $R$-ratios \cite{Grilli:1973wg,CERADINI197380,YEKUTIELI1970301}.
    \item[ii)] $\sqrt{\hat s}/\Lambda_{QCD}\gg \mathcal{O}(10)$ leads to $2\rightarrow \mathit{many}$ mesons, with the final states characterized in terms of QCD-jets rather than explicit meson multiplicities. An example is collisions at the Z-pole at LEP \cite{MALAZA1991169, Ammosov:1972cx}.
\end{itemize} 
Although the boundary between $2\to \mathit {few}$ and $2\to \mathit {many}$ regimes is not well defined, this classification helps define the final state. 
A similar framework can be used in TC-like theories. The analogue of QCD-mesons ($\pi$'s etc.) are `EW-bosons', i.e. the lightest bound states of the hyper-quarks. They are identified as (pseudo-)scalars that are not charged under QCD's $SU(3)$ and have masses around the EW scale. The `EW-bosons' include the longitudinal components of the massive gauge bosons and the Higgs, together with other spin-0 states, whose number and properties depend on the global symmetry of the specific TC-like model. If we consider a benchmark $\Lambda_{HC} = 2$~TeV, the regime $2\rightarrow \mathit{few}$ would occur if $\sqrt{\hat s}\sim 10$~TeV. On the other hand,  $\sqrt{\hat s}\gtrsim 50$~TeV would be necessary for the realization of the $2\rightarrow \mathit{many}$ regime.
While the latter is beyond the reach of any foreseeable terrestrial accelerator, the former regime with $\sqrt{\hat s}\sim 10$~TeV can be reached at the proposed $100$~TeV hadron collider (FCC-hh) \cite{FCC:2025lpp} and at a high-energy muon collider \cite{InternationalMuonCollider:2025sys}.

In this letter, we present a proof-of-principle study of $2\to {\mathit {few}}$ EW-boson process in TC-like theories at future high energy colliders. The pair production $pp (q\bar{q}')/\mu^+\mu^- \to q_{HC} \bar{q}'_{HC}$, followed by fragmentation and hadronization under $G_{HC}$, is treated as the analogue of $e^+e^- \to q\bar{q}$ in QCD at low-energy colliders. In both cases, the hadronizing quarks are produced via their EW interactions. We show that such a multi EW-boson final state yields characteristic kinematic properties, which allow for a clear distinction from purely SM processes. This signature, together with the direct discovery of light resonances (produced singly or in pairs), provides a smoking gun for the discovery of composite dynamics in the Higgs sector.

%\vspace{0.5cm}

For concreteness, we consider a strong dynamics as close as possible to that of QCD, hence $G_{HC} = SU(N)_{HC}$ with $q_{HC}$ in the fundamental representation.\footnote{Our discussion is also applicable to larger number of flavors and colors, as long as the fermion multiplicity falls below the conformal windowsill \cite{Ryttov:2007sr}. Beyond this regime, simulating the so called near-conformal theories need new developments in Pythia~\cite{Kulkarni:2025rsl}. } In the most minimal model, 
the hyper-quarks are organized into doublets under the global flavor symmetry $G_F\equiv SU(2)_L\times SU(2)_R$:
\begin{equation}
    q_{HC} \equiv \begin{pmatrix}
U \\ D
\end{pmatrix}_{L} \oplus  \begin{pmatrix}
U \\ D
\end{pmatrix}_{R}\,.
\end{equation}
The flavor symmetry is partly gauged by the EW symmetry $G_{EW} = SU(2)_L \times U(1)_Y$. These fields transform under ${SU(N)_{HC}} \times G_{EW}$ as follows:
\begin{equation*}
    \begin{pmatrix}
U \\ D
\end{pmatrix}_{L} \equiv (N,2,y), \; U_{R} \equiv (N,1, y+\frac{1}{2}), \; D_{R} \equiv (N,1,y-\frac{1}{2})\,.
\end{equation*}
At a scale $\Lambda_{HC}$, the formation of a hyper-quark condensate,
\begin{equation}
\langle \overline{U_L} U_R \rangle = \langle \overline{D_L} D_R \rangle \neq 0\,,
\end{equation}
is responsible for the breaking of the flavor symmetry to its diagonal subgroup, $G_F \to H\equiv SU(2)_V$. In this minimalist construction, the three Goldstone bosons are identified with the longitudinal components of the $W^\pm$ and $Z$ gauge bosons, while the analogue of the CP-even $\sigma$ meson can be identified with the Higgs boson \cite{Yamawaki:1985zg,Bando:1986bg,Appelquist:2010gy,Foadi:2012bb}. 
Instead of identifying individual bosons, we label all mesons as  EW-gauge bosons to remain model agnostic.
Note that, for $N=2$ ($y=0$), this model corresponds to the ultra-minimal Technicolor of \cite{Ryttov:2008xe}, which also allows to embed the Higgs as a pseudo-Goldstone boson \cite{Cacciapaglia_2014,Cacciapaglia:2020kgq} thanks to an enhanced global symmetry. Instead, $N=3$ ($y=1/6$) leads to the old-school Technicolor \cite{Susskind:1978ms,FARHI1981277}, where a light $\sigma$--Higgs can be obtained via the addition of heavy TC-quark flavors \cite{Brower:2015owo,Witzel:2020hyr}. Furthermore, top partial compositeness \cite{Kaplan:1991dc} can also be implemented by means of the same heavy flavors \cite{Vecchi:2015fma}.
Finally, $N=4$ models have also been considered within the framework of partial compositeness in \cite{Ferretti:2013kya,Ferretti:2016upr}. Models with extended global symmetries can also be constructed, leading to the presence of additional light EW-bosons \cite{Kaplan:1983sm,DUGAN1985299} that will decay predominantly into pairs of EW gauge bosons or third generation fermions \cite{Ferretti:2016upr,Cacciapaglia:2022bax}. In some models, the EW-bosons may include a dark matter candidate \cite{Frigerio:2012uc,Wu:2017iji,Cai:2019cow,Cacciapaglia:2019ixa,Cai:2020njb}, hence leading to missing energy in the decay. In the following we only consider TC-like models, where $\Lambda_{HC}$ is closely related to the EW scale via the meson decay constant $f_\pi \sim \frac{\sqrt{N}}{4\pi} \Lambda_{HC}$, with $f_\pi \equiv v = 2 m_W/g$. In models where the Higgs emerges as a pseudo-Goldstone, a hierarchy between $f_\pi$ and $v$ is necessary, with typically $v/f_\pi \lesssim 0.1$ \cite{Agashe:2005dk,Agashe:2006at}, hence leading to values of $\Lambda_{HC}$ too large to be accessible at future colliders in the $2\to few$ regime. Higgs coupling measurements to gauge bosons provide one of the strongest bound on $\Lambda_{HC}$. The existing constraints permit $\sim 3\%$ deviation with an improvement upto $1\%$ expected at HL-LHC and consistent with $\Lambda_{HC}\sim 2$ TeV. Probing larger values of $\Lambda_{HC}$ requires a sub-percent sensitivity that is expected at muon colliders and the FCC-ee/hh facilities. The supplementary material \cite{SM} (see also references \cite{Giudice_2007, BUCHALLA2015602,stefanek2025nonuniversalprobescompositehiggs, deblas2024globalsmeftfitsfuture,Giudice_2007, Banerjee_2018,PhysRevD.110.013003,atlas2025highlightshllhcphysicsprojections,mlynarikova2023higgsphysicshllhc,PhysRevD.109.073009,Chiesa_2020,PhysRevD.106.073007,FCC:2025lpp,Selvaggi:2025kmd,maura2025higgsselfcouplingfccee} therein) offers a detailed insight on the different regimes of $\Lambda_{HC}$ that can be probed by the current and future experiments.
\vspace{0.5cm}

For the collider signatures, we consider Drell-Yan production of hyper-quark pairs for both the FCC-hh and muon colliders as follows:
\begin{itemize} 
\item[A)] $p p\rightarrow q_{HC}\bar q_{HC}$ and
\item[B)]~$\mu^+\mu^- \rightarrow q_{HC}\bar q_{HC}$.
\end{itemize}
The processes are simulated using {\tt{MADGRAPH}} \cite{Alwall_2011} and {\tt{PYTHIA}} \cite{bierlich2022comprehensiveguidephysicsusage} at 100 TeV and 10 TeV center-of-mass energy, respectively. The data was generated from the {\tt{PYTHIA}} codes in \cite{cacciapaglia_2025_16424021}. The production is due to EW gauge bosons, following the quantum numbers specified above (for $y=0$). After hadronization, the outgoing EW-bosons are required to be in the central region of the detector, satisfying a pseudo-rapidity cut $|\eta|<2.5$.  In the minimal set-up, the final state bosons will be composed of the longitudinal components of the gauge bosons, $W^\pm$ and $Z$.  In the extended set-ups, other bosons in the electroweak coset $G_F/H$, will also be formed. Other heavy resonances, with masses around $\Lambda_{HC}$ will also be produced during hadronization, such as spin-1 TC-$\rho$ mesons.

Several parameters govern the properties of the signal in our model, e.g., the multiplicity, phase-space distributions of the EW-bosons. They correspond to the process of fragmentation as well as hadronization, as implemented in the {\tt PYTHIA} HV module \cite{Carloni_2010, Carloni_2011}. Among those are {\tt ProbVector}, $a_L, b_L, r_Q, \sigma_{m_q}$ -- parameters that control the hadronization -- and $\Lambda_{HC}$ together with the order of running coupling -- parameters that control fragmentation or shower \cite{HV}. We chose to investigate a subset of these parameters, which have the largest impact on the signal properties. {\tt ProbVector} controls the probability for a formed meson to be a $\rho$, as it ranges from $0$ to $1$, and its value should be fitted from data. 
As we are close to the chiral limit, $m_\pi \ll m_\rho$,  the value for this parameter cannot be easily fixed \cite{Liu:2025bbc}. 
We use the QCD template value $0.3$, 
while noting that its increase reduces the   scalar mesons multiplicity in the final state.
The decay, $\rho\to \pi \pi$  could also further enrich the pion multiplicity, however in realistic models other decay channels are also open, rendering the final state model dependent, even in the most minimal model \cite{Casalbuoni:1993su,BuarqueFranzosi:2016ooy,Caliri:2024jdk}. Hence, in our analysis, we conservatively ignore them and only count the scalar states directly produced via the hadronization.  
We also concentrate on $a_L, b_L$, which are two dimensionless parameters entering the Lund fragmentation function \cite{Andersson:1983ia,Andersson:1983jt}.
Given the absence of data for TC-like models, the values of the Lund parameters are chosen to be the same as the values recommended for QCD ($a_L= 0.3$, and $b_{L}$ = 0.087)~\cite{Liu:2025bbc}. The constituent quark mass is set to $\Lambda_{HC}$. The supplementary material \cite{SM}provides further insights on the dependence of the results on these parameters.

%the value of $b_L$ for QCD-like models is extracted from that of QCD by the expression $b_L= \frac{b^{QCD}_{L}\times (0.3)^{3}}{M^{2}_{q_{TC}}}$ ( $b_{L}$ = 0.087)~\cite{Liu:2025bbc}. \sk{We can now mention the exact parameters we use and we don't have to mention the scaling.} The constituent quark mass, $M_{q_{TC}}$ is set to be equal to $\Lambda_{TC}$.

The length of shower is controlled by the ratio $\sqrt{\hat s}/\Lambda_{HC}$ and thus it is an important parameter in determining the final-state EW-boson multiplicity. 
%%%%%%%%%%%%%%%%%%%%%%%%%%%%%%%%%%%%%%%%%%%%%%%%%%%%%%%%%%%%%%%%%%%%%%%%%%%%%%%%%%%%%%%%%%%%%%%
% adding plot for the multiplicity
\begin{figure}[htbp]   % plot of the multiplicity at two invariant masses.
   % \centering
   \begin{center}
\includegraphics[width=9.0cm,height=4.5cm]{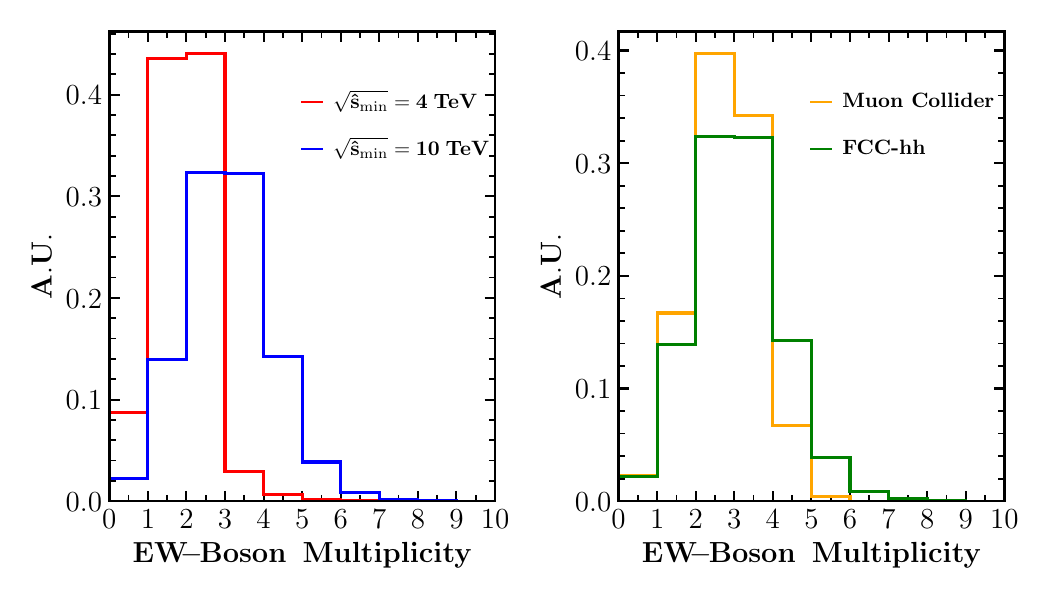}       
  \end{center}
  \caption{Final state EW-boson multiplicity.  \textbf{Left:}Distribution at FCC-hh for $\sqrt{\hat{s}_{min}} = 4$~TeV (red) and $\sqrt{\hat{s}_{min}} = 10$~TeV (blue). \textbf{ Right:} Comparison between the FCC-hh at $\sqrt{\hat{s}_{min}} = 10$~TeV (green) and a muon collider at fixed $\sqrt{\hat{s}} = 10$~TeV (orange)}
    \label{fig:mesoncount}
\end{figure}
% \begin{figure}[htbp]   % plot of the multiplicity for two different colliders
%     \centering
%     \includegraphics[width=\linewidth]{paper_new/plots/mesons_count_collider.pdf}
%     \caption{Comparison of the final state EW-boson count for the FCC-hh at $\sqrt{\hat{s}_{min}} = 10$~TeV (green) and a muon collider at fixed $\sqrt{\hat{s}} = 10$~TeV (orange).}
%     \label{fig:mesoncount_comparison}
% \end{figure}
%%%%%%%%%%%%%%%%%%%%%%%%%%%%%%%%%%%%%%%%%%%%%%%%%%%%%%%%%%%%%%%%%%%%%%%%%%%%%%%%%%%%%%%%%%%%%%%
%Fixing $\Lambda_{TC}=2$~TeV, the red(blue) curves give the meson multiplicity for a minimum \sk{partonic} collision energy $\sqrt{\hat s_{min}} = 4(10)$~TeV. 
This is illustrated in the left plot of Fig.~\ref{fig:mesoncount}, where the meson multiplicity for a minimum partonic collision energy $\sqrt{\hat s_{min}} = 4(10)$~TeV (red, blue curves) is shown for fixed $\Lambda_{HC}=2$~TeV.
As the constituent quark mass is set to $\Lambda_{HC}$, the kinematics for the Drell-Yan production automatically ensures that the minimum invariant mass is $\sqrt{\hat s_{min}}= 4$ TeV. At this point, no shower is initiated, and hence the corresponding distribution (red curve) peaks at two bosons. On the other hand, $\sqrt{\hat s_{min}} = 10$~TeV allows for a perturbative $q_{HC} \bar{q}_{HC}$ production process, followed by shower and hadronization, resulting in larger multiplicities. 
 This is reflected in the blue curve where the boson multiplicity peaks at three. The tail of the distribution as seen in Fig.~\ref{fig:mesoncount} results from variation of partonic center-of-mass energy $\sqrt{\hat{s}} > \sqrt{\hat{s}_{min}}$, in a $pp$ system. The corresponding plot for the muon collider, operating at a fixed $\sqrt{\hat{s}}$, has a significantly suppressed tail. The right plot of Fig.~\ref{fig:mesoncount} quantifies this statement by comparing the expected EW-boson multiplicity for FCC-hh and muon colliders.   

The signal, with a desired multiplicity of final state bosons, has an irreducible SM background of the form $pp(\mu\mu)\rightarrow XY, XYZ$, where $X,Y,Z\equiv$ EW bosons ($W^\pm, Z$). The higher multiplicity final states in the background are primarily due to pair production of  off-shell EW bosons and their eventual decay into multi-body final states.  The background cross sections are evaluated at leading order from {\tt{MADGRAPH}} \cite{Alwall_2011} and {\tt{PYTHIA}} \cite{bierlich2022comprehensiveguidephysicsusage} by requiring that each of the EW-boson in the final state satisfies $|\eta|<2.5$. The signal cross sections are estimated by factoring the Drell-Yan cross-section ($\sigma(pp/\mu\mu\rightarrow q_{HC}\bar q'_{HC} )$) with the corresponding efficiency for the three and four body final state.
These cross sections can be estimated for different multiplicity of final states. For illustration we consider $\Lambda_{HC}=2$. 
For three boson EW-bosons final state, signal cross-sections are,  $\sim 64$ ab at FCC-hh with $\sqrt{\hat s_{min}}/\Lambda_{HC}=5$  and $\sim 758$ ab at the muon collider with $\sqrt{\hat s}/\Lambda_{HC}=5$. The corresponding values for the background, are $\sim 422$ and $\sim 4701$ ab, respectively. The numbers for the four boson final state can also be similarly estimated and reflect similar patterns of hierarchy between the signal and the background.

\begin{figure}[htbp]
  \centering
     \includegraphics[width=9.0cm,height=4.5cm]{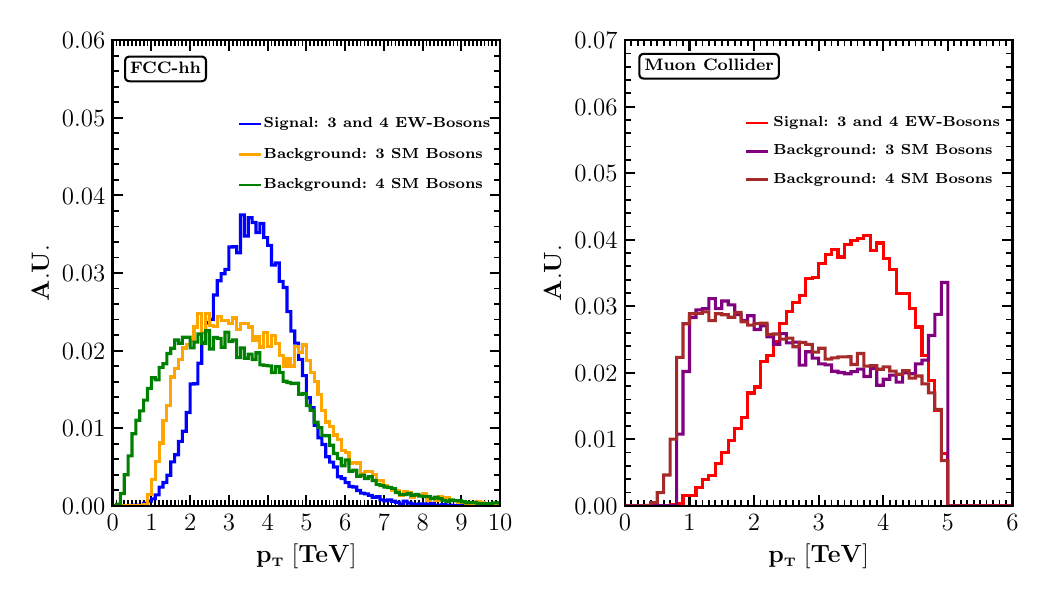} 
     \caption{Leading $p_{T}$ distribution of the EW-bosons (signal), and the three and four SM bosons for  FCC-hh(left) and  muon collider(right).}
  \label{fig:signal}
\end{figure}

%\vspace{0.5cm}
It is interesting to note that the signal and SM cross sections are roughly of the same order of magnitude. This paints a highly optimistic scenario for a potential discovery of a multi EW-boson final state due to fragmentation. This situation is further bolstered by comparing the transverse momentum ($p_T$) distribution of all the EW-bosons for both the signal, with $\Lambda_{TC}=2$ TeV, and background. Choosing  $\sqrt{\hat s_{min}}/\Lambda_{HC}=5$ for FCC-hh and $\sqrt{\hat s}/\Lambda_{HC}=5$ for the muon collider, the distribution is shown in
Fig.~\ref{fig:signal}. 
%%%%%%%%%%%%%%%%%%%%%%%%%%%%%%%%%%%%%%%%%%%%%%%%%%%%%%%%%%%%%%%%%%%%%%%%%%%%%%%%%%%%%%%%%%%%%%%%%
The top and bottom plots correspond to the FCC-hh and muon collider, respectively. All the EW-bosons from the signal have consistently higher $p_T$, as they emerge from the fragmentation of the (heavy) hyper-quarks and are predominantly produced in the transverse plane. On the other hand,  the EW-bosons from the background have a tendency to be in the forward direction, and typically only one of the total of three and four bosons in the final state is boosted. Thus, the signal event rate in the higher $p_T$ bins is expected to be at least comparable with the background. The $p_T$ distribution for the FCC-hh has a smoother longer tail as larger values of partonic center-of-mass energy is accessible. On the other hand, the distribution for the muon collider exhibits a sharp cut-off at 5 TeV as it is operates at a fixed center of mass energy of 10 TeV.

Hence, the ratio $\sqrt{\hat s}/\Lambda$ plays a crucial role in the ability to identify a signal over the background. While higher values of the ratio $\sqrt{\hat s}/\Lambda$ would lead to higher $p_T$ bosons and higher multiplicity of final states for the signal, one would be limited by a falling cross-section for the signal. As an illustration,  Table \ref{tab:crossection} shows the cross sections at FCC-hh after cuts for $\Lambda_{HC}=2$ TeV and three different values of the ratio, $\sqrt{\hat s_{min}}/\Lambda=5,10,15$. We also remark that for fixed $\sqrt{\hat s_{min}}/\Lambda$, the cross-sections decrease with increasing $\Lambda_{HC}$ and grow linearly with $N$.
\begin{table}[htb!]
    \centering
    \begin{tabular}{|l|c|c|c|l|}
    \hline
        \multirow{2}{*}{$\sqrt{\hat s_{min}}$ (TeV)}& \multicolumn{2}{c|}{3 Bosons}& \multicolumn{2}{c|}{4 Bosons}\\
         \cline{2-3} \cline{4-5}
           &$\sigma_{SM}$ (ab) &$\sigma_{sig}$ (ab) & $\sigma_{SM}$ (ab) & $\sigma_{sig}$ (ab) \\ 
\hline
10 & 422.3 & 64.7 & 96.6 & 29.0 \\  
\hline
20 & 13.8 & 0.8 & 4.8 & 1.0   \\        
\hline
30 &0.75& 0.01 & 0.4 & 0.0  \\           
\hline
    \end{tabular}
   \caption{Comparison of SM ($\sigma_{SM}$) and signal ($\sigma_{sig}$) cross sections for $\Lambda_{HC}=2$ TeV at FCC-hh. Three and four boson final state are shown at three different  $\sqrt{\hat{s}_{min}}$. }
    \label{tab:crossection}
\end{table}

The extent in $\Lambda_{HC}$ to which a given signal can be discriminated over the background can be quantified using the binned sensitivity \cite{Cowan_2011} $Z=\sqrt{\sum\limits_{i=1}^{M} \left(2(s_i+b_i)\log\left[1+\frac{s_i}{b_i}\right]-2s_i\right)}$,
% \begin{equation}
% Z=\sqrt{\sum\limits_{i=1}^{M} \left(2(s_i+b_i)\log\left[1+\frac{s_i}{b_i}\right]-2s_i\right)}\,,
% \label{eq:zvalue}
% \end{equation}
where the sum runs over the bins for a given discriminating variable and $s_i(b_i)$ is the expected number of signal (background) events in the $i^{th}$ bin at a given luminosity. As $p_T$ is identified as a useful variable to discriminate signal over background, we calculate the $Z$-score using the $p_T$ distributions with bins of size $20$ GeV over the range $[0,10]$ TeV. As the study is meant to be a proof-of-principle, the EW bosons for both the signal and background are considered as final-state objects. As a result, the $Z$ value corresponds to the maximum possible sensitivity that can be achieved for a given $\Lambda_{HC}$. 
%%%%%%%%%%%%%%%%%%%%%%%%%%%%%%%%%%%%%%%%%%%%%%%%%%%%%%%%%%%%%%%%%%%%%%%%%%%%%%%%%%%%%%%%%%%%%%%%%
\begin{figure}[htbp!]
    \centering
    \includegraphics[width=6cm]{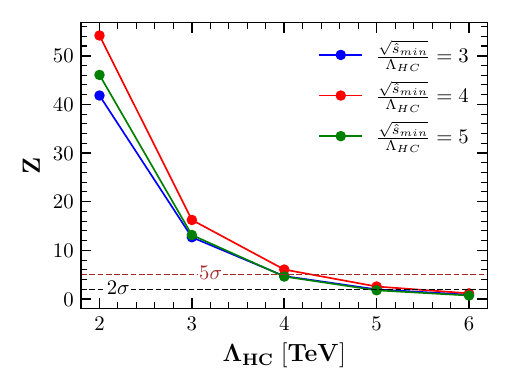}
    \caption{Variation of  $\mathit{Z}$ (signal sensitivity) with $\Lambda_{\mathit{HC}}$ for fixed $\sqrt{\hat{s}_{min}}/\Lambda_{\mathit{HC}}$ ratio.}
    \label{fig:z_value}
\end{figure} 
%%%%%%%%%%%%%%%%%%%%%%%%%%%%%%%%%%%%%%%%%%%%%%%%%%%%%%%%%%%%%%%%%%%%%%%%%%%%%%%%%%%%%%%%%%%%%%%%%
Fig.~\ref{fig:z_value} illustrates the sensitivity $Z$ for three and four EW-boson final state as a function of $\Lambda_{HC}$ at FCC-hh, for three different $\sqrt{\hat s_{min}}/\Lambda$ settings/benchmarks.  For a given value of $\Lambda_{HC}$, an interesting hierarchy in $Z$ between the three different values of $\sqrt{\hat s_{min}}/\Lambda$ emerges.  This is because $\sqrt{\hat s_{min}}/\Lambda=4$ has the maximum cross section for three and four boson final state. For $\sqrt{\hat s_{min}}/\Lambda=5$ on the other hand, a higher multiplicity of final states is preferred.
The $2\sigma$ line illustrates the extent to which a bound on $\Lambda_{HC}$ can be achieved using this methodology. The prospects for a muon collider are relatively less stringent and only $\Lambda_{HC}$ up to 3 TeV can be excluded at 95$\%$ with this method.

\begin{figure}[htbp]
  \centering
    \includegraphics[width=\linewidth]{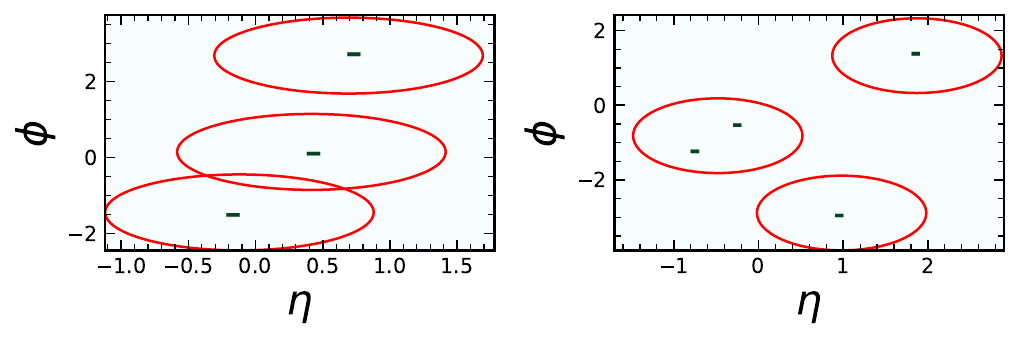} \caption{Visualization of signal in the form of jets with $R = 1$ and represented by closed red contours.}    \label{fig:eta-phi}
  \end{figure}

In a complete analysis, the identification of each boson will be performed through its decay products. As the signal EW bosons are highly boosted, their decay products will likely be concentrated in a small angular region and hence can be classified as jets. Fig.~\ref{fig:eta-phi} illustrates the distributions of EW bosons (green dots) in the $\eta-\phi$ plane for two representative events for the signal with $\Lambda_{TC}=2$ TeV. The region enclosed by the red boundary around each boson represents the circular disc with radius $R<1$. This indicates the fact that the eventual decay products of the EW bosons will be distributed over a small region and hence identified as boosted jets. The right plot is particularly interesting as it suggests the clustering of two EW bosons in a single jet, which is extremely rare in the SM. 
%%%%%%%%%%%%%%%%%%%%%%%%%%%%%%%%%%%%%%%%%%%%%%%%%%%%%%%%%%%%%%%%%%%%%%%%%%%%%%%%%%%%%%%%%%%% 
%%%%%%%%%%%%%%%%%%%%%%%%%%%%%%%%%%%%%%%%%%%%%%%%%%%%%%%%%%%%%%%%%%%%%%%%%%%%%%%%%%%%%%%%%%%
The high values of $\Lambda_{HC}$ and the available energies favor the exploitation of three and four EW boson final states.
However,  probing higher values of $\sqrt{\hat s_{min}}/\Lambda_{HC}$ may lead to higher multiplicities, particularly for relatively low values of $\Lambda_{HC}$.
For example, $\Lambda_{HC}=2$ TeV and $\sqrt{\hat s_{min}}/\Lambda_{HC}=10$,  can lead to about 25 signal events that contain five EW-boson events. Although this does not seem significant, it must be remembered that it may be associated with negligible backgrounds. Higher multiplicities of EW bosons may also derive from non-minimal models, where the cosets contain additional light scalars in addition to longitudinal $W$, $Z$, and Higgs \cite{DUGAN1985299,Bellazzini_2014,Agugliaro:2018vsu,Cacciapaglia:2022bax}, and from the decays of the $\rho$ mesons produced via hadronization. The new scalars could decay into a pair of EW gauge bosons (including the photon), hence providing a new handle into this process. A detailed analysis would crucially depend on the specific model, and it is left for future investigations.

%%%%%%%%%%%%%%%%%%%%%%%%%%%%%%%%%%%%%%%%%%%%%%%%%%%%%%%%%%%%%%%%%%%%%%%%%%%%%%%%%%%%%%%%%%%%
\vspace{0.4cm}
Learning from QCD,  we explored a novel regime for models of composite Higgs sectors, corresponding to energies a few times larger than the resonance scale $\Lambda_{HC}$. In this regime, the effects of the new strong dynamics can be described in terms of Drell-Yan production of the fundamental hyper-quarks, followed by fragmentation and hadronization. We showed that for low values of $\Lambda_{HC}$ in the few TeV range, a future 100 TeV hadron collider (FCC-hh) and a 10 TeV muon collider provide a picture similar to the one of a few GeV lepton collider for QCD. Higgs compositeness reveals itself by the production of few EW-bosons ($W$, $Z$ and Higgs), which can be kinematically distinguished from the pure standard model backgrounds. We provide a first proof-of-principle toward the discovery of this regime of TeV-scale strong interactions. This signature provides a clear proxy for the composite nature of the Higgs sector of the standard model and for the dynamical origin of the electroweak symmetry breaking, together with the further direct search for EW-resonances.

\section*{Acknowledgments}
We thank Nishita Desai for useful discussions and collaboration in the initial phase of the project. We would like to thank MITP for hosting ``The Future of Fundamental Composite Dynamics: Theory, Colliders, Cosmology, and Tools'', where many aspects of the project took shape and were discussed.
S.K. is supported by the Austrian Science Fund research teams grant STRONG-DM (FG1) and project number P 36947-N.
A.M.I. acknowledges the generous support by SERB India through project no. SRG/2022/001003.
 A.M.I. also thanks the French Institute- Embassy of France
in India for facilitating the research trip to IP2I Lyon in December 2022, where the project was first discussed.
AMI would also like to thank IP2I Lyon for the hospitality during several collaborative visits.
The research of A.K.S is supported by the Institute Fellowship from the Indian Institute of Technology Delhi, India.
%\bibliography{biblio}

\begin{thebibliography}{66}%
\makeatletter
\providecommand \@ifxundefined [1]{%
 \@ifx{#1\undefined}
}%
\providecommand \@ifnum [1]{%
 \ifnum #1\expandafter \@firstoftwo
 \else \expandafter \@secondoftwo
 \fi
}%
\providecommand \@ifx [1]{%
 \ifx #1\expandafter \@firstoftwo
 \else \expandafter \@secondoftwo
 \fi
}%
\providecommand \natexlab [1]{#1}%
\providecommand \enquote  [1]{``#1''}%
\providecommand \bibnamefont  [1]{#1}%
\providecommand \bibfnamefont [1]{#1}%
\providecommand \citenamefont [1]{#1}%
\providecommand \href@noop [0]{\@secondoftwo}%
\providecommand \href [0]{\begingroup \@sanitize@url \@href}%
\providecommand \@href[1]{\@@startlink{#1}\@@href}%
\providecommand \@@href[1]{\endgroup#1\@@endlink}%
\providecommand \@sanitize@url [0]{\catcode `\\12\catcode `\$12\catcode
  `\&12\catcode `\#12\catcode `\^12\catcode `\_12\catcode `\%12\relax}%
\providecommand \@@startlink[1]{}%
\providecommand \@@endlink[0]{}%
\providecommand \url  [0]{\begingroup\@sanitize@url \@url }%
\providecommand \@url [1]{\endgroup\@href {#1}{\urlprefix }}%
\providecommand \urlprefix  [0]{URL }%
\providecommand \Eprint [0]{\href }%
\providecommand \doibase [0]{http://dx.doi.org/}%
\providecommand \selectlanguage [0]{\@gobble}%
\providecommand \bibinfo  [0]{\@secondoftwo}%
\providecommand \bibfield  [0]{\@secondoftwo}%
\providecommand \translation [1]{[#1]}%
\providecommand \BibitemOpen [0]{}%
\providecommand \bibitemStop [0]{}%
\providecommand \bibitemNoStop [0]{.\EOS\space}%
\providecommand \EOS [0]{\spacefactor3000\relax}%
\providecommand \BibitemShut  [1]{\csname bibitem#1\endcsname}%
\let\auto@bib@innerbib\@empty
%</preamble>
\bibitem [{\citenamefont {Susskind}(1979)}]{Susskind:1978ms}%
  \BibitemOpen
  \bibfield  {author} {\bibinfo {author} {\bibfnamefont {L.}~\bibnamefont
  {Susskind}},\ }\href {\doibase 10.1103/PhysRevD.20.2619} {\bibfield
  {journal} {\bibinfo  {journal} {Phys. Rev. D}\ }\textbf {\bibinfo {volume}
  {20}},\ \bibinfo {pages} {2619} (\bibinfo {year} {1979})}\BibitemShut
  {NoStop}%
\bibitem [{\citenamefont {Farhi}\ and\ \citenamefont
  {Susskind}(1981)}]{FARHI1981277}%
  \BibitemOpen
  \bibfield  {author} {\bibinfo {author} {\bibfnamefont {E.}~\bibnamefont
  {Farhi}}\ and\ \bibinfo {author} {\bibfnamefont {L.}~\bibnamefont
  {Susskind}},\ }\href {\doibase https://doi.org/10.1016/0370-1573(81)90173-3}
  {\bibfield  {journal} {\bibinfo  {journal} {Physics Reports}\ }\textbf
  {\bibinfo {volume} {74}},\ \bibinfo {pages} {277} (\bibinfo {year}
  {1981})}\BibitemShut {NoStop}%
\bibitem [{\citenamefont {Kaplan}\ and\ \citenamefont
  {Georgi}(1984)}]{Kaplan:1983fs}%
  \BibitemOpen
  \bibfield  {author} {\bibinfo {author} {\bibfnamefont {D.~B.}\ \bibnamefont
  {Kaplan}}\ and\ \bibinfo {author} {\bibfnamefont {H.}~\bibnamefont
  {Georgi}},\ }\href {\doibase 10.1016/0370-2693(84)91177-8} {\bibfield
  {journal} {\bibinfo  {journal} {Phys. Lett. B}\ }\textbf {\bibinfo {volume}
  {136}},\ \bibinfo {pages} {183} (\bibinfo {year} {1984})}\BibitemShut
  {NoStop}%
\bibitem [{\citenamefont {Cacciapaglia}\ and\ \citenamefont
  {Sannino}(2014)}]{Cacciapaglia_2014}%
  \BibitemOpen
  \bibfield  {author} {\bibinfo {author} {\bibfnamefont {G.}~\bibnamefont
  {Cacciapaglia}}\ and\ \bibinfo {author} {\bibfnamefont {F.}~\bibnamefont
  {Sannino}},\ }\href {\doibase 10.1007/jhep04(2014)111} {\bibfield  {journal}
  {\bibinfo  {journal} {Journal of High Energy Physics}\ }\textbf {\bibinfo
  {volume} {2014}} (\bibinfo {year} {2014}),\
  10.1007/jhep04(2014)111}\BibitemShut {NoStop}%
\bibitem [{\citenamefont {Artru}\ and\ \citenamefont
  {Mennessier}(1974)}]{ARTRU197493}%
  \BibitemOpen
  \bibfield  {author} {\bibinfo {author} {\bibfnamefont {X.}~\bibnamefont
  {Artru}}\ and\ \bibinfo {author} {\bibfnamefont {G.}~\bibnamefont
  {Mennessier}},\ }\href {\doibase
  https://doi.org/10.1016/0550-3213(74)90360-5} {\bibfield  {journal} {\bibinfo
   {journal} {Nuclear Physics B}\ }\textbf {\bibinfo {volume} {70}},\ \bibinfo
  {pages} {93} (\bibinfo {year} {1974})}\BibitemShut {NoStop}%
\bibitem [{\citenamefont {Chun}\ and\ \citenamefont
  {Buchanan}(1993)}]{Chun:1992qs}%
  \BibitemOpen
  \bibfield  {author} {\bibinfo {author} {\bibfnamefont {S.~B.}\ \bibnamefont
  {Chun}}\ and\ \bibinfo {author} {\bibfnamefont {C.~D.}\ \bibnamefont
  {Buchanan}},\ }\href {\doibase 10.1016/0370-2693(93)90616-P} {\bibfield
  {journal} {\bibinfo  {journal} {Phys. Lett. B}\ }\textbf {\bibinfo {volume}
  {308}},\ \bibinfo {pages} {153} (\bibinfo {year} {1993})}\BibitemShut
  {NoStop}%
\bibitem [{\citenamefont {Grilli}\ \emph {et~al.}(1973)\citenamefont {Grilli}
  \emph {et~al.}}]{Grilli:1973wg}%
  \BibitemOpen
  \bibfield  {author} {\bibinfo {author} {\bibfnamefont {M.}~\bibnamefont
  {Grilli}} \emph {et~al.},\ }\href {\doibase 10.1007/BF02784093} {\bibfield
  {journal} {\bibinfo  {journal} {Nuovo Cim. A}\ }\textbf {\bibinfo {volume}
  {13}},\ \bibinfo {pages} {593} (\bibinfo {year} {1973})}\BibitemShut
  {NoStop}%
\bibitem [{\citenamefont {Ceradini}\ \emph {et~al.}(1973)\citenamefont
  {Ceradini}, \citenamefont {Conversi}, \citenamefont {D'Angelo}, \citenamefont
  {Paoluzi}, \citenamefont {Santonico},\ and\ \citenamefont
  {Visentin}}]{CERADINI197380}%
  \BibitemOpen
  \bibfield  {author} {\bibinfo {author} {\bibfnamefont {F.}~\bibnamefont
  {Ceradini}}, \bibinfo {author} {\bibfnamefont {M.}~\bibnamefont {Conversi}},
  \bibinfo {author} {\bibfnamefont {S.}~\bibnamefont {D'Angelo}}, \bibinfo
  {author} {\bibfnamefont {L.}~\bibnamefont {Paoluzi}}, \bibinfo {author}
  {\bibfnamefont {R.}~\bibnamefont {Santonico}}, \ and\ \bibinfo {author}
  {\bibfnamefont {R.}~\bibnamefont {Visentin}},\ }\href {\doibase
  https://doi.org/10.1016/0370-2693(73)90574-1} {\bibfield  {journal} {\bibinfo
   {journal} {Physics Letters B}\ }\textbf {\bibinfo {volume} {47}},\ \bibinfo
  {pages} {80} (\bibinfo {year} {1973})}\BibitemShut {NoStop}%
\bibitem [{\citenamefont {Yekutieli}\ \emph {et~al.}(1970)\citenamefont
  {Yekutieli}, \citenamefont {Toaff}, \citenamefont {Shapira}, \citenamefont
  {Ronat}, \citenamefont {Lyons}, \citenamefont {Eisenberg}, \citenamefont
  {Carmel}, \citenamefont {Fridman}, \citenamefont {Maurer}, \citenamefont
  {Strub}, \citenamefont {Voltolini}, \citenamefont {Cuer},\ and\ \citenamefont
  {Grunhaus}}]{YEKUTIELI1970301}%
  \BibitemOpen
  \bibfield  {author} {\bibinfo {author} {\bibfnamefont {G.}~\bibnamefont
  {Yekutieli}}, \bibinfo {author} {\bibfnamefont {S.}~\bibnamefont {Toaff}},
  \bibinfo {author} {\bibfnamefont {A.}~\bibnamefont {Shapira}}, \bibinfo
  {author} {\bibfnamefont {E.}~\bibnamefont {Ronat}}, \bibinfo {author}
  {\bibfnamefont {L.}~\bibnamefont {Lyons}}, \bibinfo {author} {\bibfnamefont
  {Y.}~\bibnamefont {Eisenberg}}, \bibinfo {author} {\bibfnamefont
  {Z.}~\bibnamefont {Carmel}}, \bibinfo {author} {\bibfnamefont
  {A.}~\bibnamefont {Fridman}}, \bibinfo {author} {\bibfnamefont
  {G.}~\bibnamefont {Maurer}}, \bibinfo {author} {\bibfnamefont
  {R.}~\bibnamefont {Strub}}, \bibinfo {author} {\bibfnamefont
  {C.}~\bibnamefont {Voltolini}}, \bibinfo {author} {\bibfnamefont
  {P.}~\bibnamefont {Cuer}}, \ and\ \bibinfo {author} {\bibfnamefont
  {J.}~\bibnamefont {Grunhaus}},\ }\href {\doibase
  https://doi.org/10.1016/0550-3213(70)90294-4} {\bibfield  {journal} {\bibinfo
   {journal} {Nuclear Physics B}\ }\textbf {\bibinfo {volume} {18}},\ \bibinfo
  {pages} {301} (\bibinfo {year} {1970})}\BibitemShut {NoStop}%
\bibitem [{\citenamefont {Malaza}\ \emph {et~al.}(1991)\citenamefont {Malaza},
  \citenamefont {Ritchie}, \citenamefont {Solms}, \citenamefont {{von
  Oertzen}},\ and\ \citenamefont {Miller}}]{MALAZA1991169}%
  \BibitemOpen
  \bibfield  {author} {\bibinfo {author} {\bibfnamefont {E.}~\bibnamefont
  {Malaza}}, \bibinfo {author} {\bibfnamefont {R.}~\bibnamefont {Ritchie}},
  \bibinfo {author} {\bibfnamefont {F.}~\bibnamefont {Solms}}, \bibinfo
  {author} {\bibfnamefont {D.}~\bibnamefont {{von Oertzen}}}, \ and\ \bibinfo
  {author} {\bibfnamefont {H.}~\bibnamefont {Miller}},\ }\href {\doibase
  https://doi.org/10.1016/0370-2693(91)90762-F} {\bibfield  {journal} {\bibinfo
   {journal} {Physics Letters B}\ }\textbf {\bibinfo {volume} {266}},\ \bibinfo
  {pages} {169} (\bibinfo {year} {1991})}\BibitemShut {NoStop}%
\bibitem [{\citenamefont {Ammosov}\ \emph {et~al.}(1972)\citenamefont {Ammosov}
  \emph {et~al.}}]{Ammosov:1972cx}%
  \BibitemOpen
  \bibfield  {author} {\bibinfo {author} {\bibfnamefont {V.~V.}\ \bibnamefont
  {Ammosov}} \emph {et~al.},\ }\href {\doibase 10.1016/0370-2693(72)90121-9}
  {\bibfield  {journal} {\bibinfo  {journal} {Phys. Lett. B}\ }\textbf
  {\bibinfo {volume} {42}},\ \bibinfo {pages} {519} (\bibinfo {year}
  {1972})}\BibitemShut {NoStop}%
\bibitem [{\citenamefont {Benedikt}\ \emph {et~al.}(2025)\citenamefont
  {Benedikt} \emph {et~al.}}]{FCC:2025lpp}%
  \BibitemOpen
  \bibfield  {author} {\bibinfo {author} {\bibfnamefont {M.}~\bibnamefont
  {Benedikt}} \emph {et~al.} (\bibinfo {collaboration} {FCC}),\ }\href
  {\doibase 10.17181/CERN.9DKX.TDH9} {\  (\bibinfo {year} {2025}),\
  10.17181/CERN.9DKX.TDH9},\ \Eprint {http://arxiv.org/abs/2505.00272}
  {arXiv:2505.00272 [hep-ex]} \BibitemShut {NoStop}%
\bibitem [{\citenamefont {Accettura}\ \emph {et~al.}(2025)\citenamefont
  {Accettura} \emph {et~al.}}]{InternationalMuonCollider:2025sys}%
  \BibitemOpen
  \bibfield  {author} {\bibinfo {author} {\bibfnamefont {C.}~\bibnamefont
  {Accettura}} \emph {et~al.} (\bibinfo {collaboration} {International Muon
  Collider}),\ }\href@noop {} {\  (\bibinfo {year} {2025})},\ \Eprint
  {http://arxiv.org/abs/2504.21417} {arXiv:2504.21417 [physics.acc-ph]}
  \BibitemShut {NoStop}%
\bibitem [{\citenamefont {Ryttov}\ and\ \citenamefont
  {Sannino}(2007)}]{Ryttov:2007sr}%
  \BibitemOpen
  \bibfield  {author} {\bibinfo {author} {\bibfnamefont {T.~A.}\ \bibnamefont
  {Ryttov}}\ and\ \bibinfo {author} {\bibfnamefont {F.}~\bibnamefont
  {Sannino}},\ }\href {\doibase 10.1103/PhysRevD.76.105004} {\bibfield
  {journal} {\bibinfo  {journal} {Phys. Rev. D}\ }\textbf {\bibinfo {volume}
  {76}},\ \bibinfo {pages} {105004} (\bibinfo {year} {2007})},\ \Eprint
  {http://arxiv.org/abs/0707.3166} {arXiv:0707.3166 [hep-th]} \BibitemShut
  {NoStop}%
\bibitem [{\citenamefont {Kulkarni}\ \emph {et~al.}(2025)\citenamefont
  {Kulkarni}, \citenamefont {Lockyer},\ and\ \citenamefont
  {Strassler}}]{Kulkarni:2025rsl}%
  \BibitemOpen
  \bibfield  {author} {\bibinfo {author} {\bibfnamefont {S.}~\bibnamefont
  {Kulkarni}}, \bibinfo {author} {\bibfnamefont {J.}~\bibnamefont {Lockyer}}, \
  and\ \bibinfo {author} {\bibfnamefont {M.~J.}\ \bibnamefont {Strassler}},\
  }\href@noop {} {\  (\bibinfo {year} {2025})},\ \Eprint
  {http://arxiv.org/abs/2502.18566} {arXiv:2502.18566 [hep-ph]} \BibitemShut
  {NoStop}%
\bibitem [{\citenamefont {Yamawaki}\ \emph {et~al.}(1986)\citenamefont
  {Yamawaki}, \citenamefont {Bando},\ and\ \citenamefont
  {Matumoto}}]{Yamawaki:1985zg}%
  \BibitemOpen
  \bibfield  {author} {\bibinfo {author} {\bibfnamefont {K.}~\bibnamefont
  {Yamawaki}}, \bibinfo {author} {\bibfnamefont {M.}~\bibnamefont {Bando}}, \
  and\ \bibinfo {author} {\bibfnamefont {K.-i.}\ \bibnamefont {Matumoto}},\
  }\href {\doibase 10.1103/PhysRevLett.56.1335} {\bibfield  {journal} {\bibinfo
   {journal} {Phys. Rev. Lett.}\ }\textbf {\bibinfo {volume} {56}},\ \bibinfo
  {pages} {1335} (\bibinfo {year} {1986})}\BibitemShut {NoStop}%
\bibitem [{\citenamefont {Bando}\ \emph {et~al.}(1986)\citenamefont {Bando},
  \citenamefont {Matumoto},\ and\ \citenamefont {Yamawaki}}]{Bando:1986bg}%
  \BibitemOpen
  \bibfield  {author} {\bibinfo {author} {\bibfnamefont {M.}~\bibnamefont
  {Bando}}, \bibinfo {author} {\bibfnamefont {K.-i.}\ \bibnamefont {Matumoto}},
  \ and\ \bibinfo {author} {\bibfnamefont {K.}~\bibnamefont {Yamawaki}},\
  }\href {\doibase 10.1016/0370-2693(86)91516-9} {\bibfield  {journal}
  {\bibinfo  {journal} {Phys. Lett. B}\ }\textbf {\bibinfo {volume} {178}},\
  \bibinfo {pages} {308} (\bibinfo {year} {1986})}\BibitemShut {NoStop}%
\bibitem [{\citenamefont {Appelquist}\ and\ \citenamefont
  {Bai}(2010)}]{Appelquist:2010gy}%
  \BibitemOpen
  \bibfield  {author} {\bibinfo {author} {\bibfnamefont {T.}~\bibnamefont
  {Appelquist}}\ and\ \bibinfo {author} {\bibfnamefont {Y.}~\bibnamefont
  {Bai}},\ }\href {\doibase 10.1103/PhysRevD.82.071701} {\bibfield  {journal}
  {\bibinfo  {journal} {Phys. Rev. D}\ }\textbf {\bibinfo {volume} {82}},\
  \bibinfo {pages} {071701} (\bibinfo {year} {2010})},\ \Eprint
  {http://arxiv.org/abs/1006.4375} {arXiv:1006.4375 [hep-ph]} \BibitemShut
  {NoStop}%
\bibitem [{\citenamefont {Foadi}\ \emph {et~al.}(2013)\citenamefont {Foadi},
  \citenamefont {Frandsen},\ and\ \citenamefont {Sannino}}]{Foadi:2012bb}%
  \BibitemOpen
  \bibfield  {author} {\bibinfo {author} {\bibfnamefont {R.}~\bibnamefont
  {Foadi}}, \bibinfo {author} {\bibfnamefont {M.~T.}\ \bibnamefont {Frandsen}},
  \ and\ \bibinfo {author} {\bibfnamefont {F.}~\bibnamefont {Sannino}},\ }\href
  {\doibase 10.1103/PhysRevD.87.095001} {\bibfield  {journal} {\bibinfo
  {journal} {Phys. Rev. D}\ }\textbf {\bibinfo {volume} {87}},\ \bibinfo
  {pages} {095001} (\bibinfo {year} {2013})},\ \Eprint
  {http://arxiv.org/abs/1211.1083} {arXiv:1211.1083 [hep-ph]} \BibitemShut
  {NoStop}%
\bibitem [{\citenamefont {Ryttov}\ and\ \citenamefont
  {Sannino}(2008)}]{Ryttov:2008xe}%
  \BibitemOpen
  \bibfield  {author} {\bibinfo {author} {\bibfnamefont {T.~A.}\ \bibnamefont
  {Ryttov}}\ and\ \bibinfo {author} {\bibfnamefont {F.}~\bibnamefont
  {Sannino}},\ }\href {\doibase 10.1103/PhysRevD.78.115010} {\bibfield
  {journal} {\bibinfo  {journal} {Phys. Rev. D}\ }\textbf {\bibinfo {volume}
  {78}},\ \bibinfo {pages} {115010} (\bibinfo {year} {2008})},\ \Eprint
  {http://arxiv.org/abs/0809.0713} {arXiv:0809.0713 [hep-ph]} \BibitemShut
  {NoStop}%
\bibitem [{\citenamefont {Cacciapaglia}\ \emph {et~al.}(2020)\citenamefont
  {Cacciapaglia}, \citenamefont {Pica},\ and\ \citenamefont
  {Sannino}}]{Cacciapaglia:2020kgq}%
  \BibitemOpen
  \bibfield  {author} {\bibinfo {author} {\bibfnamefont {G.}~\bibnamefont
  {Cacciapaglia}}, \bibinfo {author} {\bibfnamefont {C.}~\bibnamefont {Pica}},
  \ and\ \bibinfo {author} {\bibfnamefont {F.}~\bibnamefont {Sannino}},\ }\href
  {\doibase 10.1016/j.physrep.2020.07.002} {\bibfield  {journal} {\bibinfo
  {journal} {Phys. Rept.}\ }\textbf {\bibinfo {volume} {877}},\ \bibinfo
  {pages} {1} (\bibinfo {year} {2020})},\ \Eprint
  {http://arxiv.org/abs/2002.04914} {arXiv:2002.04914 [hep-ph]} \BibitemShut
  {NoStop}%
\bibitem [{\citenamefont {Brower}\ \emph {et~al.}(2016)\citenamefont {Brower},
  \citenamefont {Hasenfratz}, \citenamefont {Rebbi}, \citenamefont {Weinberg},\
  and\ \citenamefont {Witzel}}]{Brower:2015owo}%
  \BibitemOpen
  \bibfield  {author} {\bibinfo {author} {\bibfnamefont {R.~C.}\ \bibnamefont
  {Brower}}, \bibinfo {author} {\bibfnamefont {A.}~\bibnamefont {Hasenfratz}},
  \bibinfo {author} {\bibfnamefont {C.}~\bibnamefont {Rebbi}}, \bibinfo
  {author} {\bibfnamefont {E.}~\bibnamefont {Weinberg}}, \ and\ \bibinfo
  {author} {\bibfnamefont {O.}~\bibnamefont {Witzel}},\ }\href {\doibase
  10.1103/PhysRevD.93.075028} {\bibfield  {journal} {\bibinfo  {journal} {Phys.
  Rev. D}\ }\textbf {\bibinfo {volume} {93}},\ \bibinfo {pages} {075028}
  (\bibinfo {year} {2016})},\ \Eprint {http://arxiv.org/abs/1512.02576}
  {arXiv:1512.02576 [hep-ph]} \BibitemShut {NoStop}%
\bibitem [{\citenamefont {Witzel}\ \emph {et~al.}(2021)\citenamefont {Witzel},
  \citenamefont {Hasenfratz},\ and\ \citenamefont {Peterson}}]{Witzel:2020hyr}%
  \BibitemOpen
  \bibfield  {author} {\bibinfo {author} {\bibfnamefont {O.}~\bibnamefont
  {Witzel}}, \bibinfo {author} {\bibfnamefont {A.}~\bibnamefont {Hasenfratz}},
  \ and\ \bibinfo {author} {\bibfnamefont {C.~T.}\ \bibnamefont {Peterson}}
  (\bibinfo {collaboration} {Lattice Strong Dynamics}),\ }\href {\doibase
  10.22323/1.390.0675} {\bibfield  {journal} {\bibinfo  {journal} {PoS}\
  }\textbf {\bibinfo {volume} {ICHEP2020}},\ \bibinfo {pages} {675} (\bibinfo
  {year} {2021})},\ \Eprint {http://arxiv.org/abs/2011.05175} {arXiv:2011.05175
  [hep-ph]} \BibitemShut {NoStop}%
\bibitem [{\citenamefont {Kaplan}(1991)}]{Kaplan:1991dc}%
  \BibitemOpen
  \bibfield  {author} {\bibinfo {author} {\bibfnamefont {D.~B.}\ \bibnamefont
  {Kaplan}},\ }\href {\doibase 10.1016/S0550-3213(05)80021-5} {\bibfield
  {journal} {\bibinfo  {journal} {Nucl. Phys. B}\ }\textbf {\bibinfo {volume}
  {365}},\ \bibinfo {pages} {259} (\bibinfo {year} {1991})}\BibitemShut
  {NoStop}%
\bibitem [{\citenamefont {Vecchi}(2017)}]{Vecchi:2015fma}%
  \BibitemOpen
  \bibfield  {author} {\bibinfo {author} {\bibfnamefont {L.}~\bibnamefont
  {Vecchi}},\ }\href {\doibase 10.1007/JHEP02(2017)094} {\bibfield  {journal}
  {\bibinfo  {journal} {JHEP}\ }\textbf {\bibinfo {volume} {02}},\ \bibinfo
  {pages} {094} (\bibinfo {year} {2017})},\ \Eprint
  {http://arxiv.org/abs/1506.00623} {arXiv:1506.00623 [hep-ph]} \BibitemShut
  {NoStop}%
\bibitem [{\citenamefont {Ferretti}\ and\ \citenamefont
  {Karateev}(2014)}]{Ferretti:2013kya}%
  \BibitemOpen
  \bibfield  {author} {\bibinfo {author} {\bibfnamefont {G.}~\bibnamefont
  {Ferretti}}\ and\ \bibinfo {author} {\bibfnamefont {D.}~\bibnamefont
  {Karateev}},\ }\href {\doibase 10.1007/JHEP03(2014)077} {\bibfield  {journal}
  {\bibinfo  {journal} {JHEP}\ }\textbf {\bibinfo {volume} {03}},\ \bibinfo
  {pages} {077} (\bibinfo {year} {2014})},\ \Eprint
  {http://arxiv.org/abs/1312.5330} {arXiv:1312.5330 [hep-ph]} \BibitemShut
  {NoStop}%
\bibitem [{\citenamefont {Ferretti}(2016)}]{Ferretti:2016upr}%
  \BibitemOpen
  \bibfield  {author} {\bibinfo {author} {\bibfnamefont {G.}~\bibnamefont
  {Ferretti}},\ }\href {\doibase 10.1007/JHEP06(2016)107} {\bibfield  {journal}
  {\bibinfo  {journal} {JHEP}\ }\textbf {\bibinfo {volume} {06}},\ \bibinfo
  {pages} {107} (\bibinfo {year} {2016})},\ \Eprint
  {http://arxiv.org/abs/1604.06467} {arXiv:1604.06467 [hep-ph]} \BibitemShut
  {NoStop}%
\bibitem [{\citenamefont {Kaplan}\ \emph {et~al.}(1984)\citenamefont {Kaplan},
  \citenamefont {Georgi},\ and\ \citenamefont {Dimopoulos}}]{Kaplan:1983sm}%
  \BibitemOpen
  \bibfield  {author} {\bibinfo {author} {\bibfnamefont {D.~B.}\ \bibnamefont
  {Kaplan}}, \bibinfo {author} {\bibfnamefont {H.}~\bibnamefont {Georgi}}, \
  and\ \bibinfo {author} {\bibfnamefont {S.}~\bibnamefont {Dimopoulos}},\
  }\href {\doibase 10.1016/0370-2693(84)91178-X} {\bibfield  {journal}
  {\bibinfo  {journal} {Phys. Lett. B}\ }\textbf {\bibinfo {volume} {136}},\
  \bibinfo {pages} {187} (\bibinfo {year} {1984})}\BibitemShut {NoStop}%
\bibitem [{\citenamefont {Dugan}\ \emph {et~al.}(1985)\citenamefont {Dugan},
  \citenamefont {Georgi},\ and\ \citenamefont {Kaplan}}]{DUGAN1985299}%
  \BibitemOpen
  \bibfield  {author} {\bibinfo {author} {\bibfnamefont {M.~J.}\ \bibnamefont
  {Dugan}}, \bibinfo {author} {\bibfnamefont {H.}~\bibnamefont {Georgi}}, \
  and\ \bibinfo {author} {\bibfnamefont {D.~B.}\ \bibnamefont {Kaplan}},\
  }\href {\doibase https://doi.org/10.1016/0550-3213(85)90221-4} {\bibfield
  {journal} {\bibinfo  {journal} {Nuclear Physics B}\ }\textbf {\bibinfo
  {volume} {254}},\ \bibinfo {pages} {299} (\bibinfo {year}
  {1985})}\BibitemShut {NoStop}%
\bibitem [{\citenamefont {Cacciapaglia}\ \emph {et~al.}(2022)\citenamefont
  {Cacciapaglia}, \citenamefont {Flacke}, \citenamefont {Kunkel}, \citenamefont
  {Porod},\ and\ \citenamefont {Schwarze}}]{Cacciapaglia:2022bax}%
  \BibitemOpen
  \bibfield  {author} {\bibinfo {author} {\bibfnamefont {G.}~\bibnamefont
  {Cacciapaglia}}, \bibinfo {author} {\bibfnamefont {T.}~\bibnamefont
  {Flacke}}, \bibinfo {author} {\bibfnamefont {M.}~\bibnamefont {Kunkel}},
  \bibinfo {author} {\bibfnamefont {W.}~\bibnamefont {Porod}}, \ and\ \bibinfo
  {author} {\bibfnamefont {L.}~\bibnamefont {Schwarze}},\ }\href {\doibase
  10.1007/JHEP12(2022)087} {\bibfield  {journal} {\bibinfo  {journal} {JHEP}\
  }\textbf {\bibinfo {volume} {12}},\ \bibinfo {pages} {087} (\bibinfo {year}
  {2022})},\ \Eprint {http://arxiv.org/abs/2210.01826} {arXiv:2210.01826
  [hep-ph]} \BibitemShut {NoStop}%
\bibitem [{\citenamefont {Frigerio}\ \emph {et~al.}(2012)\citenamefont
  {Frigerio}, \citenamefont {Pomarol}, \citenamefont {Riva},\ and\
  \citenamefont {Urbano}}]{Frigerio:2012uc}%
  \BibitemOpen
  \bibfield  {author} {\bibinfo {author} {\bibfnamefont {M.}~\bibnamefont
  {Frigerio}}, \bibinfo {author} {\bibfnamefont {A.}~\bibnamefont {Pomarol}},
  \bibinfo {author} {\bibfnamefont {F.}~\bibnamefont {Riva}}, \ and\ \bibinfo
  {author} {\bibfnamefont {A.}~\bibnamefont {Urbano}},\ }\href {\doibase
  10.1007/JHEP07(2012)015} {\bibfield  {journal} {\bibinfo  {journal} {JHEP}\
  }\textbf {\bibinfo {volume} {07}},\ \bibinfo {pages} {015} (\bibinfo {year}
  {2012})},\ \Eprint {http://arxiv.org/abs/1204.2808} {arXiv:1204.2808
  [hep-ph]} \BibitemShut {NoStop}%
\bibitem [{\citenamefont {Wu}\ \emph {et~al.}(2017)\citenamefont {Wu},
  \citenamefont {Ma}, \citenamefont {Zhang},\ and\ \citenamefont
  {Cacciapaglia}}]{Wu:2017iji}%
  \BibitemOpen
  \bibfield  {author} {\bibinfo {author} {\bibfnamefont {Y.}~\bibnamefont
  {Wu}}, \bibinfo {author} {\bibfnamefont {T.}~\bibnamefont {Ma}}, \bibinfo
  {author} {\bibfnamefont {B.}~\bibnamefont {Zhang}}, \ and\ \bibinfo {author}
  {\bibfnamefont {G.}~\bibnamefont {Cacciapaglia}},\ }\href {\doibase
  10.1007/JHEP11(2017)058} {\bibfield  {journal} {\bibinfo  {journal} {JHEP}\
  }\textbf {\bibinfo {volume} {11}},\ \bibinfo {pages} {058} (\bibinfo {year}
  {2017})},\ \Eprint {http://arxiv.org/abs/1703.06903} {arXiv:1703.06903
  [hep-ph]} \BibitemShut {NoStop}%
\bibitem [{\citenamefont {Cai}\ \emph {et~al.}(2020)\citenamefont {Cai},
  \citenamefont {Zhang}, \citenamefont {Cacciapaglia}, \citenamefont
  {Rosenlyst},\ and\ \citenamefont {Frandsen}}]{Cai:2019cow}%
  \BibitemOpen
  \bibfield  {author} {\bibinfo {author} {\bibfnamefont {C.}~\bibnamefont
  {Cai}}, \bibinfo {author} {\bibfnamefont {H.-H.}\ \bibnamefont {Zhang}},
  \bibinfo {author} {\bibfnamefont {G.}~\bibnamefont {Cacciapaglia}}, \bibinfo
  {author} {\bibfnamefont {M.}~\bibnamefont {Rosenlyst}}, \ and\ \bibinfo
  {author} {\bibfnamefont {M.~T.}\ \bibnamefont {Frandsen}},\ }\href {\doibase
  10.1103/PhysRevLett.125.021801} {\bibfield  {journal} {\bibinfo  {journal}
  {Phys. Rev. Lett.}\ }\textbf {\bibinfo {volume} {125}},\ \bibinfo {pages}
  {021801} (\bibinfo {year} {2020})},\ \Eprint
  {http://arxiv.org/abs/1911.12130} {arXiv:1911.12130 [hep-ph]} \BibitemShut
  {NoStop}%
\bibitem [{\citenamefont {Cacciapaglia}\ \emph {et~al.}(2019)\citenamefont
  {Cacciapaglia}, \citenamefont {Cai}, \citenamefont {Deandrea},\ and\
  \citenamefont {Kushwaha}}]{Cacciapaglia:2019ixa}%
  \BibitemOpen
  \bibfield  {author} {\bibinfo {author} {\bibfnamefont {G.}~\bibnamefont
  {Cacciapaglia}}, \bibinfo {author} {\bibfnamefont {H.}~\bibnamefont {Cai}},
  \bibinfo {author} {\bibfnamefont {A.}~\bibnamefont {Deandrea}}, \ and\
  \bibinfo {author} {\bibfnamefont {A.}~\bibnamefont {Kushwaha}},\ }\href
  {\doibase 10.1007/JHEP10(2019)035} {\bibfield  {journal} {\bibinfo  {journal}
  {JHEP}\ }\textbf {\bibinfo {volume} {10}},\ \bibinfo {pages} {035} (\bibinfo
  {year} {2019})},\ \Eprint {http://arxiv.org/abs/1904.09301} {arXiv:1904.09301
  [hep-ph]} \BibitemShut {NoStop}%
\bibitem [{\citenamefont {Cai}\ and\ \citenamefont
  {Cacciapaglia}(2021)}]{Cai:2020njb}%
  \BibitemOpen
  \bibfield  {author} {\bibinfo {author} {\bibfnamefont {H.}~\bibnamefont
  {Cai}}\ and\ \bibinfo {author} {\bibfnamefont {G.}~\bibnamefont
  {Cacciapaglia}},\ }\href {\doibase 10.1103/PhysRevD.103.055002} {\bibfield
  {journal} {\bibinfo  {journal} {Phys. Rev. D}\ }\textbf {\bibinfo {volume}
  {103}},\ \bibinfo {pages} {055002} (\bibinfo {year} {2021})},\ \Eprint
  {http://arxiv.org/abs/2007.04338} {arXiv:2007.04338 [hep-ph]} \BibitemShut
  {NoStop}%
\bibitem [{\citenamefont {Agashe}\ and\ \citenamefont
  {Contino}(2006)}]{Agashe:2005dk}%
  \BibitemOpen
  \bibfield  {author} {\bibinfo {author} {\bibfnamefont {K.}~\bibnamefont
  {Agashe}}\ and\ \bibinfo {author} {\bibfnamefont {R.}~\bibnamefont
  {Contino}},\ }\href {\doibase 10.1016/j.nuclphysb.2006.02.011} {\bibfield
  {journal} {\bibinfo  {journal} {Nucl. Phys. B}\ }\textbf {\bibinfo {volume}
  {742}},\ \bibinfo {pages} {59} (\bibinfo {year} {2006})},\ \Eprint
  {http://arxiv.org/abs/hep-ph/0510164} {arXiv:hep-ph/0510164} \BibitemShut
  {NoStop}%
\bibitem [{\citenamefont {Agashe}\ \emph {et~al.}(2006)\citenamefont {Agashe},
  \citenamefont {Contino}, \citenamefont {Da~Rold},\ and\ \citenamefont
  {Pomarol}}]{Agashe:2006at}%
  \BibitemOpen
  \bibfield  {author} {\bibinfo {author} {\bibfnamefont {K.}~\bibnamefont
  {Agashe}}, \bibinfo {author} {\bibfnamefont {R.}~\bibnamefont {Contino}},
  \bibinfo {author} {\bibfnamefont {L.}~\bibnamefont {Da~Rold}}, \ and\
  \bibinfo {author} {\bibfnamefont {A.}~\bibnamefont {Pomarol}},\ }\href
  {\doibase 10.1016/j.physletb.2006.08.005} {\bibfield  {journal} {\bibinfo
  {journal} {Phys. Lett. B}\ }\textbf {\bibinfo {volume} {641}},\ \bibinfo
  {pages} {62} (\bibinfo {year} {2006})},\ \Eprint
  {http://arxiv.org/abs/hep-ph/0605341} {arXiv:hep-ph/0605341} \BibitemShut
  {NoStop}%
\bibitem [{SM()}]{SM}%
  \BibitemOpen
  \href@noop {} {}\bibinfo {note} {See Supplemental Material at
  \url{https://link.aps.org/supplemental/10.1103/fdjh-kjy4} for the Lund
  parameters, probvector, and precision couplings on $\Lambda_{HC}$, which
  includes references \cite{Giudice_2007,
  BUCHALLA2015602,stefanek2025nonuniversalprobescompositehiggs,
  deblas2024globalsmeftfitsfuture,Giudice_2007,
  Banerjee_2018,PhysRevD.110.013003,atlas2025highlightshllhcphysicsprojections,mlynarikova2023higgsphysicshllhc,PhysRevD.109.073009,Chiesa_2020,PhysRevD.106.073007,FCC:2025lpp,Selvaggi:2025kmd,maura2025higgsselfcouplingfccee}.}\BibitemShut
  {Stop}%
\bibitem [{\citenamefont {Giudice}\ \emph {et~al.}(2007)\citenamefont
  {Giudice}, \citenamefont {Grojean}, \citenamefont {Pomarol},\ and\
  \citenamefont {Rattazzi}}]{Giudice_2007}%
  \BibitemOpen
  \bibfield  {author} {\bibinfo {author} {\bibfnamefont {G.~F.}\ \bibnamefont
  {Giudice}}, \bibinfo {author} {\bibfnamefont {C.}~\bibnamefont {Grojean}},
  \bibinfo {author} {\bibfnamefont {A.}~\bibnamefont {Pomarol}}, \ and\
  \bibinfo {author} {\bibfnamefont {R.}~\bibnamefont {Rattazzi}},\ }\href
  {\doibase 10.1088/1126-6708/2007/06/045} {\bibfield  {journal} {\bibinfo
  {journal} {Journal of High Energy Physics}\ }\textbf {\bibinfo {volume}
  {2007}},\ \bibinfo {pages} {045–045} (\bibinfo {year} {2007})}\BibitemShut
  {NoStop}%
\bibitem [{\citenamefont {Buchalla}\ \emph {et~al.}(2015)\citenamefont
  {Buchalla}, \citenamefont {Catà},\ and\ \citenamefont
  {Krause}}]{BUCHALLA2015602}%
  \BibitemOpen
  \bibfield  {author} {\bibinfo {author} {\bibfnamefont {G.}~\bibnamefont
  {Buchalla}}, \bibinfo {author} {\bibfnamefont {O.}~\bibnamefont {Catà}}, \
  and\ \bibinfo {author} {\bibfnamefont {C.}~\bibnamefont {Krause}},\ }\href
  {\doibase https://doi.org/10.1016/j.nuclphysb.2015.03.024} {\bibfield
  {journal} {\bibinfo  {journal} {Nuclear Physics B}\ }\textbf {\bibinfo
  {volume} {894}},\ \bibinfo {pages} {602} (\bibinfo {year}
  {2015})}\BibitemShut {NoStop}%
\bibitem [{\citenamefont
  {Stefanek}(2025)}]{stefanek2025nonuniversalprobescompositehiggs}%
  \BibitemOpen
  \bibfield  {author} {\bibinfo {author} {\bibfnamefont {B.~A.}\ \bibnamefont
  {Stefanek}},\ }\href {https://arxiv.org/abs/2407.09593} {\enquote {\bibinfo
  {title} {Non-universal probes of composite higgs models: New bounds and
  prospects for fcc-ee},}\ } (\bibinfo {year} {2025}),\ \Eprint
  {http://arxiv.org/abs/2407.09593} {arXiv:2407.09593 [hep-ph]} \BibitemShut
  {NoStop}%
\bibitem [{\citenamefont {de~Blas}\ \emph {et~al.}(2024)\citenamefont
  {de~Blas}, \citenamefont {Du}, \citenamefont {Grojean}, \citenamefont {Gu},
  \citenamefont {Miralles}, \citenamefont {Peskin}, \citenamefont {Tian},
  \citenamefont {Vos},\ and\ \citenamefont
  {Vryonidou}}]{deblas2024globalsmeftfitsfuture}%
  \BibitemOpen
  \bibfield  {author} {\bibinfo {author} {\bibfnamefont {J.}~\bibnamefont
  {de~Blas}}, \bibinfo {author} {\bibfnamefont {Y.}~\bibnamefont {Du}},
  \bibinfo {author} {\bibfnamefont {C.}~\bibnamefont {Grojean}}, \bibinfo
  {author} {\bibfnamefont {J.}~\bibnamefont {Gu}}, \bibinfo {author}
  {\bibfnamefont {V.}~\bibnamefont {Miralles}}, \bibinfo {author}
  {\bibfnamefont {M.~E.}\ \bibnamefont {Peskin}}, \bibinfo {author}
  {\bibfnamefont {J.}~\bibnamefont {Tian}}, \bibinfo {author} {\bibfnamefont
  {M.}~\bibnamefont {Vos}}, \ and\ \bibinfo {author} {\bibfnamefont
  {E.}~\bibnamefont {Vryonidou}},\ }\href {https://arxiv.org/abs/2206.08326}
  {\enquote {\bibinfo {title} {Global smeft fits at future colliders},}\ }
  (\bibinfo {year} {2024}),\ \Eprint {http://arxiv.org/abs/2206.08326}
  {arXiv:2206.08326 [hep-ph]} \BibitemShut {NoStop}%
\bibitem [{\citenamefont {Banerjee}\ \emph {et~al.}(2018)\citenamefont
  {Banerjee}, \citenamefont {Bhattacharyya}, \citenamefont {Kumar},\ and\
  \citenamefont {Ray}}]{Banerjee_2018}%
  \BibitemOpen
  \bibfield  {author} {\bibinfo {author} {\bibfnamefont {A.}~\bibnamefont
  {Banerjee}}, \bibinfo {author} {\bibfnamefont {G.}~\bibnamefont
  {Bhattacharyya}}, \bibinfo {author} {\bibfnamefont {N.}~\bibnamefont
  {Kumar}}, \ and\ \bibinfo {author} {\bibfnamefont {T.~S.}\ \bibnamefont
  {Ray}},\ }\href {\doibase 10.1007/jhep03(2018)062} {\bibfield  {journal}
  {\bibinfo  {journal} {Journal of High Energy Physics}\ }\textbf {\bibinfo
  {volume} {2018}} (\bibinfo {year} {2018}),\
  10.1007/jhep03(2018)062}\BibitemShut {NoStop}%
\bibitem [{\citenamefont {Heo}\ \emph {et~al.}(2024)\citenamefont {Heo},
  \citenamefont {Jung},\ and\ \citenamefont {Lee}}]{PhysRevD.110.013003}%
  \BibitemOpen
  \bibfield  {author} {\bibinfo {author} {\bibfnamefont {Y.}~\bibnamefont
  {Heo}}, \bibinfo {author} {\bibfnamefont {D.-W.}\ \bibnamefont {Jung}}, \
  and\ \bibinfo {author} {\bibfnamefont {J.~S.}\ \bibnamefont {Lee}},\ }\href
  {\doibase 10.1103/PhysRevD.110.013003} {\bibfield  {journal} {\bibinfo
  {journal} {Phys. Rev. D}\ }\textbf {\bibinfo {volume} {110}},\ \bibinfo
  {pages} {013003} (\bibinfo {year} {2024})}\BibitemShut {NoStop}%
\bibitem [{\citenamefont {ATLAS}\ and\ \citenamefont
  {Collaborations}(2025)}]{atlas2025highlightshllhcphysicsprojections}%
  \BibitemOpen
  \bibfield  {author} {\bibinfo {author} {\bibnamefont {ATLAS}}\ and\ \bibinfo
  {author} {\bibfnamefont {C.}~\bibnamefont {Collaborations}},\ }\href
  {https://arxiv.org/abs/2504.00672} {\enquote {\bibinfo {title} {Highlights of
  the hl-lhc physics projections by atlas and cms},}\ } (\bibinfo {year}
  {2025}),\ \Eprint {http://arxiv.org/abs/2504.00672} {arXiv:2504.00672
  [hep-ex]} \BibitemShut {NoStop}%
\bibitem [{\citenamefont
  {Mlynarikova}(2023)}]{mlynarikova2023higgsphysicshllhc}%
  \BibitemOpen
  \bibfield  {author} {\bibinfo {author} {\bibfnamefont {M.}~\bibnamefont
  {Mlynarikova}},\ }\href {https://arxiv.org/abs/2307.07772} {\enquote
  {\bibinfo {title} {Higgs physics at hl-lhc},}\ } (\bibinfo {year} {2023}),\
  \Eprint {http://arxiv.org/abs/2307.07772} {arXiv:2307.07772 [hep-ex]}
  \BibitemShut {NoStop}%
\bibitem [{\citenamefont {Li}\ \emph {et~al.}(2024)\citenamefont {Li},
  \citenamefont {Liu},\ and\ \citenamefont {Lyu}}]{PhysRevD.109.073009}%
  \BibitemOpen
  \bibfield  {author} {\bibinfo {author} {\bibfnamefont {P.}~\bibnamefont
  {Li}}, \bibinfo {author} {\bibfnamefont {Z.}~\bibnamefont {Liu}}, \ and\
  \bibinfo {author} {\bibfnamefont {K.-F.}\ \bibnamefont {Lyu}},\ }\href
  {\doibase 10.1103/PhysRevD.109.073009} {\bibfield  {journal} {\bibinfo
  {journal} {Phys. Rev. D}\ }\textbf {\bibinfo {volume} {109}},\ \bibinfo
  {pages} {073009} (\bibinfo {year} {2024})}\BibitemShut {NoStop}%
\bibitem [{\citenamefont {Chiesa}\ \emph {et~al.}(2020)\citenamefont {Chiesa},
  \citenamefont {Maltoni}, \citenamefont {Mantani}, \citenamefont {Mele},
  \citenamefont {Piccinini},\ and\ \citenamefont {Zhao}}]{Chiesa_2020}%
  \BibitemOpen
  \bibfield  {author} {\bibinfo {author} {\bibfnamefont {M.}~\bibnamefont
  {Chiesa}}, \bibinfo {author} {\bibfnamefont {F.}~\bibnamefont {Maltoni}},
  \bibinfo {author} {\bibfnamefont {L.}~\bibnamefont {Mantani}}, \bibinfo
  {author} {\bibfnamefont {B.}~\bibnamefont {Mele}}, \bibinfo {author}
  {\bibfnamefont {F.}~\bibnamefont {Piccinini}}, \ and\ \bibinfo {author}
  {\bibfnamefont {X.}~\bibnamefont {Zhao}},\ }\href {\doibase
  10.1007/jhep09(2020)098} {\bibfield  {journal} {\bibinfo  {journal} {Journal
  of High Energy Physics}\ }\textbf {\bibinfo {volume} {2020}} (\bibinfo {year}
  {2020}),\ 10.1007/jhep09(2020)098}\BibitemShut {NoStop}%
\bibitem [{\citenamefont {de~Blas}\ \emph {et~al.}(2022)\citenamefont
  {de~Blas}, \citenamefont {Gu},\ and\ \citenamefont
  {Liu}}]{PhysRevD.106.073007}%
  \BibitemOpen
  \bibfield  {author} {\bibinfo {author} {\bibfnamefont {J.}~\bibnamefont
  {de~Blas}}, \bibinfo {author} {\bibfnamefont {J.}~\bibnamefont {Gu}}, \ and\
  \bibinfo {author} {\bibfnamefont {Z.}~\bibnamefont {Liu}},\ }\href {\doibase
  10.1103/PhysRevD.106.073007} {\bibfield  {journal} {\bibinfo  {journal}
  {Phys. Rev. D}\ }\textbf {\bibinfo {volume} {106}},\ \bibinfo {pages}
  {073007} (\bibinfo {year} {2022})}\BibitemShut {NoStop}%
\bibitem [{\citenamefont {Selvaggi}\ \emph {et~al.}(2025)\citenamefont
  {Selvaggi}, \citenamefont {Blondel},\ and\ \citenamefont
  {Eysermans}}]{Selvaggi:2025kmd}%
  \BibitemOpen
  \bibfield  {author} {\bibinfo {author} {\bibfnamefont {M.}~\bibnamefont
  {Selvaggi}}, \bibinfo {author} {\bibfnamefont {A.}~\bibnamefont {Blondel}}, \
  and\ \bibinfo {author} {\bibfnamefont {J.}~\bibnamefont {Eysermans}}
  (\bibinfo {collaboration} {FCC}),\ }\href {\doibase 10.17181/n78xk-qcv56} {\
  (\bibinfo {year} {2025}),\ 10.17181/n78xk-qcv56}\BibitemShut {NoStop}%
\bibitem [{\citenamefont {Maura}\ \emph {et~al.}(2025)\citenamefont {Maura},
  \citenamefont {Stefanek},\ and\ \citenamefont
  {You}}]{maura2025higgsselfcouplingfccee}%
  \BibitemOpen
  \bibfield  {author} {\bibinfo {author} {\bibfnamefont {V.}~\bibnamefont
  {Maura}}, \bibinfo {author} {\bibfnamefont {B.~A.}\ \bibnamefont {Stefanek}},
  \ and\ \bibinfo {author} {\bibfnamefont {T.}~\bibnamefont {You}},\ }\href
  {https://arxiv.org/abs/2503.13719} {\enquote {\bibinfo {title} {The higgs
  self-coupling at fcc-ee},}\ } (\bibinfo {year} {2025}),\ \Eprint
  {http://arxiv.org/abs/2503.13719} {arXiv:2503.13719 [hep-ph]} \BibitemShut
  {NoStop}%
\bibitem [{\citenamefont {Alwall}\ \emph {et~al.}(2011)\citenamefont {Alwall},
  \citenamefont {Herquet}, \citenamefont {Maltoni}, \citenamefont {Mattelaer},\
  and\ \citenamefont {Stelzer}}]{Alwall_2011}%
  \BibitemOpen
  \bibfield  {author} {\bibinfo {author} {\bibfnamefont {J.}~\bibnamefont
  {Alwall}}, \bibinfo {author} {\bibfnamefont {M.}~\bibnamefont {Herquet}},
  \bibinfo {author} {\bibfnamefont {F.}~\bibnamefont {Maltoni}}, \bibinfo
  {author} {\bibfnamefont {O.}~\bibnamefont {Mattelaer}}, \ and\ \bibinfo
  {author} {\bibfnamefont {T.}~\bibnamefont {Stelzer}},\ }\href {\doibase
  10.1007/jhep06(2011)128} {\bibfield  {journal} {\bibinfo  {journal} {Journal
  of High Energy Physics}\ }\textbf {\bibinfo {volume} {2011}} (\bibinfo {year}
  {2011}),\ 10.1007/jhep06(2011)128}\BibitemShut {NoStop}%
\bibitem [{\citenamefont {Bierlich}\ \emph {et~al.}(2022)\citenamefont
  {Bierlich}, \citenamefont {Chakraborty}, \citenamefont {Desai}, \citenamefont
  {Gellersen}, \citenamefont {Helenius}, \citenamefont {Ilten}, \citenamefont
  {Lönnblad}, \citenamefont {Mrenna}, \citenamefont {Prestel}, \citenamefont
  {Preuss}, \citenamefont {Sjöstrand}, \citenamefont {Skands}, \citenamefont
  {Utheim},\ and\ \citenamefont
  {Verheyen}}]{bierlich2022comprehensiveguidephysicsusage}%
  \BibitemOpen
  \bibfield  {author} {\bibinfo {author} {\bibfnamefont {C.}~\bibnamefont
  {Bierlich}}, \bibinfo {author} {\bibfnamefont {S.}~\bibnamefont
  {Chakraborty}}, \bibinfo {author} {\bibfnamefont {N.}~\bibnamefont {Desai}},
  \bibinfo {author} {\bibfnamefont {L.}~\bibnamefont {Gellersen}}, \bibinfo
  {author} {\bibfnamefont {I.}~\bibnamefont {Helenius}}, \bibinfo {author}
  {\bibfnamefont {P.}~\bibnamefont {Ilten}}, \bibinfo {author} {\bibfnamefont
  {L.}~\bibnamefont {Lönnblad}}, \bibinfo {author} {\bibfnamefont
  {S.}~\bibnamefont {Mrenna}}, \bibinfo {author} {\bibfnamefont
  {S.}~\bibnamefont {Prestel}}, \bibinfo {author} {\bibfnamefont {C.~T.}\
  \bibnamefont {Preuss}}, \bibinfo {author} {\bibfnamefont {T.}~\bibnamefont
  {Sjöstrand}}, \bibinfo {author} {\bibfnamefont {P.}~\bibnamefont {Skands}},
  \bibinfo {author} {\bibfnamefont {M.}~\bibnamefont {Utheim}}, \ and\ \bibinfo
  {author} {\bibfnamefont {R.}~\bibnamefont {Verheyen}},\ }\href
  {https://arxiv.org/abs/2203.11601} {\enquote {\bibinfo {title} {A
  comprehensive guide to the physics and usage of pythia 8.3},}\ } (\bibinfo
  {year} {2022}),\ \Eprint {http://arxiv.org/abs/2203.11601} {arXiv:2203.11601
  [hep-ph]} \BibitemShut {NoStop}%
\bibitem [{\citenamefont {Cacciapaglia}\ \emph {et~al.}(2025)\citenamefont
  {Cacciapaglia}, \citenamefont {Deandrea}, \citenamefont {Iyer}, \citenamefont
  {Kulkarni},\ and\ \citenamefont {Singh}}]{cacciapaglia_2025_16424021}%
  \BibitemOpen
  \bibfield  {author} {\bibinfo {author} {\bibfnamefont {G.}~\bibnamefont
  {Cacciapaglia}}, \bibinfo {author} {\bibfnamefont {A.}~\bibnamefont
  {Deandrea}}, \bibinfo {author} {\bibfnamefont {A.}~\bibnamefont {Iyer}},
  \bibinfo {author} {\bibfnamefont {S.}~\bibnamefont {Kulkarni}}, \ and\
  \bibinfo {author} {\bibfnamefont {A.~K.}\ \bibnamefont {Singh}},\ }\href
  {\doibase 10.5281/zenodo.16424021} {\enquote {\bibinfo {title} {Simulation
  code for multi-boson splashes at future colliders from electroweak
  compositeness},}\ } (\bibinfo {year} {2025})\BibitemShut {NoStop}%
\bibitem [{\citenamefont {Carloni}\ and\ \citenamefont
  {Sjöstrand}(2010)}]{Carloni_2010}%
  \BibitemOpen
  \bibfield  {author} {\bibinfo {author} {\bibfnamefont {L.}~\bibnamefont
  {Carloni}}\ and\ \bibinfo {author} {\bibfnamefont {T.}~\bibnamefont
  {Sjöstrand}},\ }\href {\doibase 10.1007/jhep09(2010)105} {\bibfield
  {journal} {\bibinfo  {journal} {Journal of High Energy Physics}\ }\textbf
  {\bibinfo {volume} {2010}} (\bibinfo {year} {2010}),\
  10.1007/jhep09(2010)105}\BibitemShut {NoStop}%
\bibitem [{\citenamefont {Carloni}\ \emph {et~al.}(2011)\citenamefont
  {Carloni}, \citenamefont {Rathsman},\ and\ \citenamefont
  {Sjöstrand}}]{Carloni_2011}%
  \BibitemOpen
  \bibfield  {author} {\bibinfo {author} {\bibfnamefont {L.}~\bibnamefont
  {Carloni}}, \bibinfo {author} {\bibfnamefont {J.}~\bibnamefont {Rathsman}}, \
  and\ \bibinfo {author} {\bibfnamefont {T.}~\bibnamefont {Sjöstrand}},\
  }\href {\doibase 10.1007/jhep04(2011)091} {\bibfield  {journal} {\bibinfo
  {journal} {Journal of High Energy Physics}\ }\textbf {\bibinfo {volume}
  {2011}} (\bibinfo {year} {2011}),\ 10.1007/jhep04(2011)091}\BibitemShut
  {NoStop}%
\bibitem [{\citenamefont {Pythia}()}]{HV}%
  \BibitemOpen
  \bibfield  {author} {\bibinfo {author} {\bibnamefont {Pythia}},\ }\href
  {https://pythia.org/latest-manual/HiddenValleyProcesses.html} {\enquote
  {\bibinfo {title} {Hidden valley processes},}\ }\BibitemShut {NoStop}%
\bibitem [{\citenamefont {Liu}\ \emph {et~al.}(2025)\citenamefont {Liu},
  \citenamefont {Lockyer},\ and\ \citenamefont {Kulkarni}}]{Liu:2025bbc}%
  \BibitemOpen
  \bibfield  {author} {\bibinfo {author} {\bibfnamefont {W.}~\bibnamefont
  {Liu}}, \bibinfo {author} {\bibfnamefont {J.}~\bibnamefont {Lockyer}}, \ and\
  \bibinfo {author} {\bibfnamefont {S.}~\bibnamefont {Kulkarni}},\ }\href@noop
  {} {\  (\bibinfo {year} {2025})},\ \Eprint {http://arxiv.org/abs/2505.03058}
  {arXiv:2505.03058 [hep-ph]} \BibitemShut {NoStop}%
\bibitem [{\citenamefont {Casalbuoni}\ \emph {et~al.}(1993)\citenamefont
  {Casalbuoni}, \citenamefont {Chiappetta}, \citenamefont {Deandrea},
  \citenamefont {De~Curtis}, \citenamefont {Dominici},\ and\ \citenamefont
  {Gatto}}]{Casalbuoni:1993su}%
  \BibitemOpen
  \bibfield  {author} {\bibinfo {author} {\bibfnamefont {R.}~\bibnamefont
  {Casalbuoni}}, \bibinfo {author} {\bibfnamefont {P.}~\bibnamefont
  {Chiappetta}}, \bibinfo {author} {\bibfnamefont {A.}~\bibnamefont
  {Deandrea}}, \bibinfo {author} {\bibfnamefont {S.}~\bibnamefont {De~Curtis}},
  \bibinfo {author} {\bibfnamefont {D.}~\bibnamefont {Dominici}}, \ and\
  \bibinfo {author} {\bibfnamefont {R.}~\bibnamefont {Gatto}},\ }\href
  {\doibase 10.1007/BF01474629} {\bibfield  {journal} {\bibinfo  {journal} {Z.
  Phys. C}\ }\textbf {\bibinfo {volume} {60}},\ \bibinfo {pages} {315}
  (\bibinfo {year} {1993})},\ \Eprint {http://arxiv.org/abs/hep-ph/9303201}
  {arXiv:hep-ph/9303201} \BibitemShut {NoStop}%
\bibitem [{\citenamefont {Buarque~Franzosi}\ \emph {et~al.}(2016)\citenamefont
  {Buarque~Franzosi}, \citenamefont {Cacciapaglia}, \citenamefont {Cai},
  \citenamefont {Deandrea},\ and\ \citenamefont
  {Frandsen}}]{BuarqueFranzosi:2016ooy}%
  \BibitemOpen
  \bibfield  {author} {\bibinfo {author} {\bibfnamefont {D.}~\bibnamefont
  {Buarque~Franzosi}}, \bibinfo {author} {\bibfnamefont {G.}~\bibnamefont
  {Cacciapaglia}}, \bibinfo {author} {\bibfnamefont {H.}~\bibnamefont {Cai}},
  \bibinfo {author} {\bibfnamefont {A.}~\bibnamefont {Deandrea}}, \ and\
  \bibinfo {author} {\bibfnamefont {M.}~\bibnamefont {Frandsen}},\ }\href
  {\doibase 10.1007/JHEP11(2016)076} {\bibfield  {journal} {\bibinfo  {journal}
  {JHEP}\ }\textbf {\bibinfo {volume} {11}},\ \bibinfo {pages} {076} (\bibinfo
  {year} {2016})},\ \Eprint {http://arxiv.org/abs/1605.01363} {arXiv:1605.01363
  [hep-ph]} \BibitemShut {NoStop}%
\bibitem [{\citenamefont {Caliri}\ \emph {et~al.}(2025)\citenamefont {Caliri},
  \citenamefont {Hadlik}, \citenamefont {Kunkel}, \citenamefont {Porod},
  \citenamefont {Verollet},\ and\ \citenamefont {Verollet}}]{Caliri:2024jdk}%
  \BibitemOpen
  \bibfield  {author} {\bibinfo {author} {\bibfnamefont {R.}~\bibnamefont
  {Caliri}}, \bibinfo {author} {\bibfnamefont {J.}~\bibnamefont {Hadlik}},
  \bibinfo {author} {\bibfnamefont {M.}~\bibnamefont {Kunkel}}, \bibinfo
  {author} {\bibfnamefont {W.}~\bibnamefont {Porod}}, \bibinfo {author}
  {\bibfnamefont {C.}~\bibnamefont {Verollet}}, \ and\ \bibinfo {author}
  {\bibfnamefont {C.}~\bibnamefont {Verollet}},\ }\href {\doibase
  10.1007/JHEP04(2025)160} {\bibfield  {journal} {\bibinfo  {journal} {JHEP}\
  }\textbf {\bibinfo {volume} {04}},\ \bibinfo {pages} {160} (\bibinfo {year}
  {2025})},\ \Eprint {http://arxiv.org/abs/2412.08720} {arXiv:2412.08720
  [hep-ph]} \BibitemShut {NoStop}%
\bibitem [{\citenamefont {Andersson}\ \emph
  {et~al.}(1983{\natexlab{a}})\citenamefont {Andersson}, \citenamefont
  {Gustafson}, \citenamefont {Ingelman},\ and\ \citenamefont
  {Sjostrand}}]{Andersson:1983ia}%
  \BibitemOpen
  \bibfield  {author} {\bibinfo {author} {\bibfnamefont {B.}~\bibnamefont
  {Andersson}}, \bibinfo {author} {\bibfnamefont {G.}~\bibnamefont
  {Gustafson}}, \bibinfo {author} {\bibfnamefont {G.}~\bibnamefont {Ingelman}},
  \ and\ \bibinfo {author} {\bibfnamefont {T.}~\bibnamefont {Sjostrand}},\
  }\href {\doibase 10.1016/0370-1573(83)90080-7} {\bibfield  {journal}
  {\bibinfo  {journal} {Phys. Rept.}\ }\textbf {\bibinfo {volume} {97}},\
  \bibinfo {pages} {31} (\bibinfo {year} {1983}{\natexlab{a}})}\BibitemShut
  {NoStop}%
\bibitem [{\citenamefont {Andersson}\ \emph
  {et~al.}(1983{\natexlab{b}})\citenamefont {Andersson}, \citenamefont
  {Gustafson},\ and\ \citenamefont {Soderberg}}]{Andersson:1983jt}%
  \BibitemOpen
  \bibfield  {author} {\bibinfo {author} {\bibfnamefont {B.}~\bibnamefont
  {Andersson}}, \bibinfo {author} {\bibfnamefont {G.}~\bibnamefont
  {Gustafson}}, \ and\ \bibinfo {author} {\bibfnamefont {B.}~\bibnamefont
  {Soderberg}},\ }\href {\doibase 10.1007/BF01407824} {\bibfield  {journal}
  {\bibinfo  {journal} {Z. Phys. C}\ }\textbf {\bibinfo {volume} {20}},\
  \bibinfo {pages} {317} (\bibinfo {year} {1983}{\natexlab{b}})}\BibitemShut
  {NoStop}%
\bibitem [{\citenamefont {Cowan}\ \emph {et~al.}(2011)\citenamefont {Cowan},
  \citenamefont {Cranmer}, \citenamefont {Gross},\ and\ \citenamefont
  {Vitells}}]{Cowan_2011}%
  \BibitemOpen
  \bibfield  {author} {\bibinfo {author} {\bibfnamefont {G.}~\bibnamefont
  {Cowan}}, \bibinfo {author} {\bibfnamefont {K.}~\bibnamefont {Cranmer}},
  \bibinfo {author} {\bibfnamefont {E.}~\bibnamefont {Gross}}, \ and\ \bibinfo
  {author} {\bibfnamefont {O.}~\bibnamefont {Vitells}},\ }\href {\doibase
  10.1140/epjc/s10052-011-1554-0} {\bibfield  {journal} {\bibinfo  {journal}
  {The European Physical Journal C}\ }\textbf {\bibinfo {volume} {71}}
  (\bibinfo {year} {2011}),\ 10.1140/epjc/s10052-011-1554-0}\BibitemShut
  {NoStop}%
\bibitem [{\citenamefont {Bellazzini}\ \emph {et~al.}(2014)\citenamefont
  {Bellazzini}, \citenamefont {Csáki},\ and\ \citenamefont
  {Serra}}]{Bellazzini_2014}%
  \BibitemOpen
  \bibfield  {author} {\bibinfo {author} {\bibfnamefont {B.}~\bibnamefont
  {Bellazzini}}, \bibinfo {author} {\bibfnamefont {C.}~\bibnamefont {Csáki}},
  \ and\ \bibinfo {author} {\bibfnamefont {J.}~\bibnamefont {Serra}},\ }\href
  {\doibase 10.1140/epjc/s10052-014-2766-x} {\bibfield  {journal} {\bibinfo
  {journal} {The European Physical Journal C}\ }\textbf {\bibinfo {volume}
  {74}} (\bibinfo {year} {2014}),\ 10.1140/epjc/s10052-014-2766-x}\BibitemShut
  {NoStop}%
\bibitem [{\citenamefont {Agugliaro}\ \emph {et~al.}(2019)\citenamefont
  {Agugliaro}, \citenamefont {Cacciapaglia}, \citenamefont {Deandrea},\ and\
  \citenamefont {De~Curtis}}]{Agugliaro:2018vsu}%
  \BibitemOpen
  \bibfield  {author} {\bibinfo {author} {\bibfnamefont {A.}~\bibnamefont
  {Agugliaro}}, \bibinfo {author} {\bibfnamefont {G.}~\bibnamefont
  {Cacciapaglia}}, \bibinfo {author} {\bibfnamefont {A.}~\bibnamefont
  {Deandrea}}, \ and\ \bibinfo {author} {\bibfnamefont {S.}~\bibnamefont
  {De~Curtis}},\ }\href {\doibase 10.1007/JHEP02(2019)089} {\bibfield
  {journal} {\bibinfo  {journal} {JHEP}\ }\textbf {\bibinfo {volume} {02}},\
  \bibinfo {pages} {089} (\bibinfo {year} {2019})},\ \Eprint
  {http://arxiv.org/abs/1808.10175} {arXiv:1808.10175 [hep-ph]} \BibitemShut
  {NoStop}%
\end{thebibliography}

\end{document}